\documentclass{article}
\usepackage{arxiv}
\usepackage{graphicx}
\usepackage[numbers,sort&compress]{natbib}
\usepackage{amsmath,amssymb,amsfonts,amsthm}
\usepackage{algorithmic}
\usepackage{enumerate}
\usepackage{mathrsfs}
\usepackage{algorithm}
\usepackage{arydshln}
\usepackage{textcomp}
\usepackage[hidelinks]{hyperref}
\hypersetup{pdftitle={Adaptive Robust Output Regulation of Uncertain Linear Time-Varying Systems},pdfauthor={Jinmeng Zha and Zhen Zhang}}
\date{}
\newtheorem{assumption}{Assumption}
\newtheorem{definition}{Definition}[section]
\newtheorem{theorem}{Theorem}[section]
\newtheorem{proposition}{Proposition}[section]
\newtheorem{remark}{Remark}[section]
\newtheorem{lemma}{Lemma}[section]
\newtheorem{corollary}{Corollary}[section]
\newtheorem{example}{Example}

\begin{document}

\title{Adaptive Robust Output Regulation of Uncertain Linear Time-Varying Systems}
\author{Jinmeng Zha and Zhen Zhang
\thanks{Corresponding author: Zhen Zhang (e-mail: zzhang@tsinghua.edu.cn).
The authors are both with the State Key Laboratory of Tribology in Advanced Equipment, Tsinghua University, Beijing 100084, China,
and Zhen Zhang is also with Beijing Key Laboratory of Transformative High-end Manufacturing Equipment and Technology, Department of Mechanical Engineering, Tsinghua University, Beijing 100084, China.}}
\maketitle

\begin{abstract}
This paper studies adaptive robust output regulation of linear time-varying systems with unknown constant parameters in both the plant and exosystem.
The design is based on nonmonic differential relations for the ideal input.
For general time-varying and periodic systems, we derive sufficient conditions for the existence of nonmonic differential relations and construct them from the original system coefficients.
Based on these relations, an adaptive controller is constructed using stable filters and a parameterized readout.
Under suitable conditions on the filter and readout, the controller achieves exact regulation for finitely separable relation coefficients, including the affine parametric uncertainty treated in Byrnes--Isidori form.
For coefficient families satisfying the approximation conditions, refining the approximation achieves any prescribed ultimate regulation accuracy with a fixed stabilization design, yielding practical regulation.
Neither persistent excitation nor parameter convergence is required.
Two numerical examples illustrate the regulation performance of the proposed controller.
Overall, the framework includes earlier constructions based on finite separation of the ideal input or monic linear differential relations with finitely separable coefficients, and achieves exact robust regulation for a broader class of uncertain systems.
\end{abstract}


\section{Introduction}\label{sec:introduction}
Output regulation is a fundamental problem in control theory.
It concerns asymptotic tracking of reference signals and/or rejection of disturbances generated by an exosystem while maintaining closed-loop stability.
The classical theory for linear time-invariant (LTI) systems was established in~\cite{davison1976robust,francis1976internal,francis1977linear} and subsequently extended to nonlinear systems~\cite{isidori1990output,jiang2025output,han2025iss,li2025global} and multi-agent systems~\cite{lin2025direct,gul2026multi,li2026data}.
For linear time-varying (LTV) systems, the problem was introduced in~\cite{pinzoni1993output}, followed by foundational results~\cite{zhang2006linear,ichikawa2006output} and further developments in internal-model construction and stabilization~\cite{sun2009trajectory,zhang2010novel,zhang2014discrete,song2015low}.

Internal-model design provides a systematic way to achieve regulation for LTI systems with model uncertainty~\cite{isidori2003robust}.
In LTV systems, however, the ideal-input mapping depends on time, and the effect of uncertainty cannot generally be described by changes in constant matrices or a fixed set of frequencies.
Despite this difficulty, robust regulation results are available for uncertain periodic systems~\cite{zhang2009tac}, linear hybrid systems with periodic jumps~\cite{marconi2013internal,carnevale2017robust}, and general time-varying systems~\cite{zha2026robust}.
Following a common LTI design approach that treats plant and exosystem uncertainties separately,~\cite{zhang2009tac} uses a robust linear internal model to accommodate plant uncertainty and introduces adaptation for unknown exosystem parameters.
In~\cite{zha2026robust}, asymptotic reproduction of the ideal input by a fixed finite-dimensional linear internal model in zero-input operation requires finite separation modulo decaying terms.
However, this property can fail even for plants with finite-dimensional affine parameter uncertainty.
Achieving exact robust regulation for a broader class of uncertain systems therefore calls for internal models beyond this fixed finite-dimensional linear structure.

One way to move beyond fixed finite-dimensional linear internal models is to use infinite-dimensional controllers.
For periodic systems,~\cite{paunonen2017robust} uses a lifting technique to construct infinite-dimensional controllers for exact robust regulation and finite-dimensional discrete-time controllers for approximate robust regulation.
To pursue exact regulation with a finite-dimensional controller, we adopt a nonlinear design based on adaptation.
Nonlinear output regulation suggests focusing directly on the ideal input required to maintain zero regulation error, since this is the signal that the internal model is designed to reproduce asymptotically~\cite{isidori1990output}.
In nonlinear systems, the effects of the plant and exosystem on this input are generally coupled~\cite{bin2022robustness}.
Since time variation can produce similarly complicated dependence on uncertainty~\cite{zha2026robust}, we study the generation of the ideal input directly and incorporate unknown plant and exosystem parameters in the same design.

Two lines of internal-model design are particularly relevant to generating the ideal input.
One starts from a differential relation between the ideal input and finitely many of its derivatives.
When the nonlinear relation is known, it can be used directly to construct an internal model~\cite{byrnes2004nonlinear}.
For unknown relations,~\cite{forte2017robust,bin2020class} use derivative prediction models and online identification, allowing prediction residuals.
For uncertain periodic and time-varying systems, the observer-form constructions in~\cite{zhang2009tac,yang2010framework} assume monic linear differential relations with finitely separable coefficients and then achieve regulation through adaptation.
The other line uses a generic internal model consisting of a suitably chosen stable linear filter and an unknown nonlinear readout~\cite{marconi2007output}.
Continuous-time identification and data-driven methods are then used to approximate the readout~\cite{bernard2020adaptive,bin2020approximate,gentilini2023data}.
The resulting ultimate regulation error is related to the prediction error of the identifier.
In adaptive implementations of both lines, the unknown derivative relation or readout must be represented by a suitable model class, whose structure is generally not explicit in the original system description.
To obtain a tractable model structure,~\cite{han2025filter,wang2024nonparametric} exploit LTI exosystems and ideal steady-state inputs that are polynomial in the exosystem state.
The resulting constant-coefficient differential relations lead to linear regressions for identification or explicit nonlinear readouts.
These developments motivate a search for an ideal-input generation relation that can be established from the original system and admits a finite-dimensional adaptive implementation.

This paper addresses adaptive output regulation of LTV systems with unknown constant parameters in both the plant and exosystem.
The main contributions are as follows.
\begin{enumerate}
\item Starting from a nonmonic differential relation for the ideal input, we construct stable filters with a parameterized readout and establish closed-loop boundedness and robust regulation without requiring persistent excitation or parameter convergence.
An exact regulator is obtained in the absence of persistent model residuals, while bounded persistent residuals yield an ultimate error bound.
The framework includes exact regulation based on fixed finite-dimensional linear internal models with decaying remainders and on monic linear differential relations with finitely separable coefficients.
Its scope is strictly broader, since exact regulation is also achieved for ideal-input families outside both classes.
\item Sufficient conditions for the existence of a nonmonic differential relation are derived for general time-varying and periodic systems, together with a finite-order construction from the original system coefficients.
The construction preserves finite separability and makes the ideal-input condition verifiable from the system description.
\item Building on the above results, constructive designs for exact and practical regulation are developed under the respective realization conditions.
The exact design applies to affine parametric uncertainty in Byrnes--Isidori form, while coefficient approximation, including Taylor expansions for analytic parameter dependence, achieves any prescribed ultimate regulation accuracy with a fixed stabilization gain, thus yielding practical regulation.
\end{enumerate}

\section{Preliminaries}\label{sec:problem}
\subsection{Problem Formulation}\label{sec:problem-formulation}
Consider the linear time-varying (LTV) system
\begin{equation}\label{eq:plant}
\begin{aligned}
\dot w&=S(t,\mu)w,\\
\dot x&=A(t,\mu)x+B(t,\mu)u+P(t,\mu)w,\\
e&=C(t,\mu)x+Q(t,\mu)w,
\end{aligned}
\end{equation}
with plant state $x\in\mathbb R^n$, exosystem state $w\in\mathbb R^q$, control input $u\in\mathbb R$, regulation error $e\in\mathbb R$, and an unknown constant parameter vector $\mu$ in a known compact set $\mathcal P\subset\mathbb R^{n_\mu}$.
Throughout, all prescribed coefficient functions and individual time channels, together with their time derivatives needed in the finite constructions below, are continuous and uniformly bounded.
The bounds are uniform for $t\ge0$ and over the indicated parameter sets, and may depend on a fixed realization index $K$ unless stated otherwise.
Sufficient derivative orders are collected in Supplementary~S9.1\footnote{The supplementary material is provided as \texttt{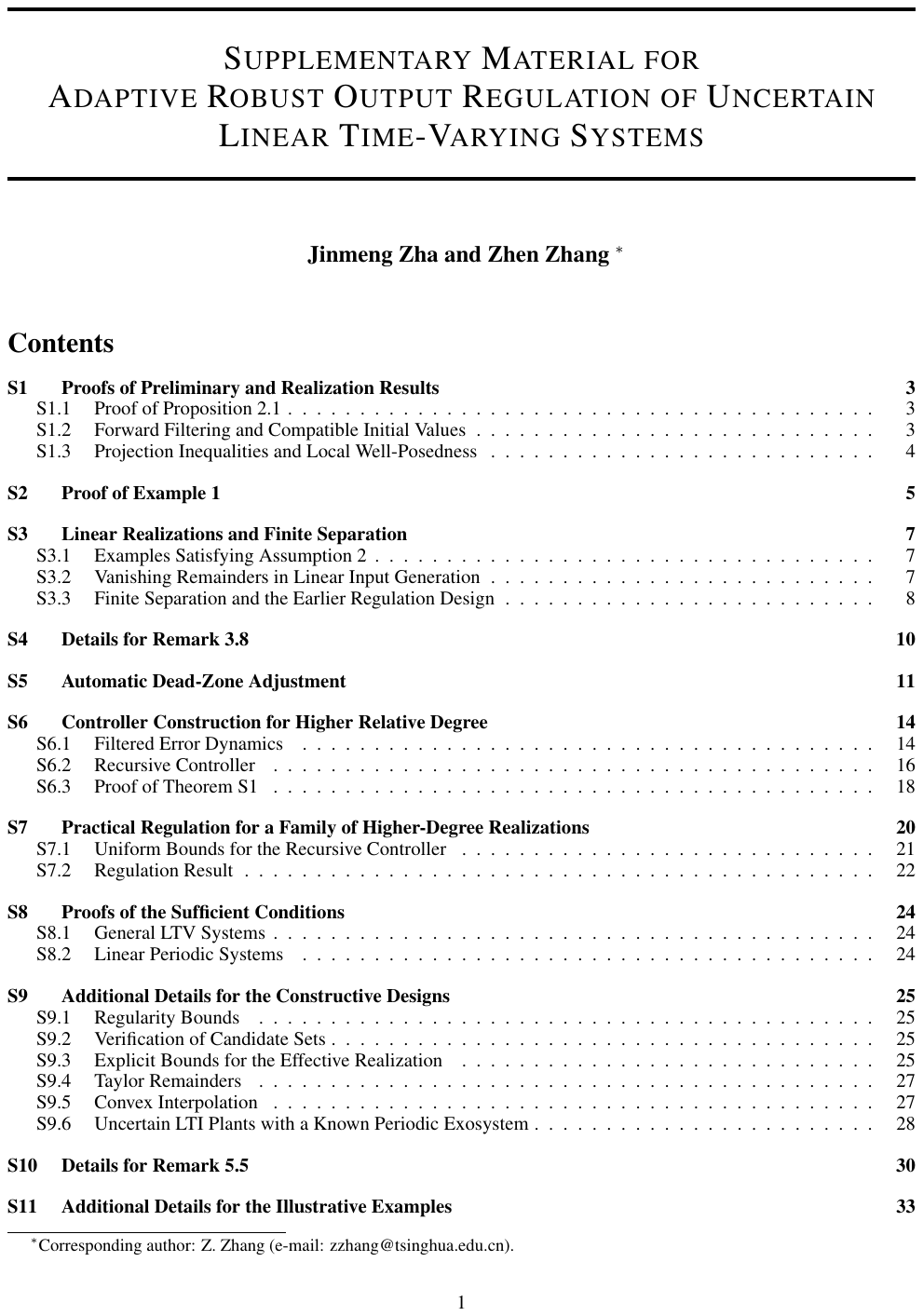} in the TeX source archive available from the arXiv page for this paper.}.
A dot and a superscript $(j)$ denote the first and $j$th time derivatives, respectively.

The robust regulation problem is to find an error-feedback time-varying controller, allowing nonlinear adaptation, such that, for every $\mu\in\mathcal P$ and initial states in prescribed bounded sets, all closed-loop solutions exist for $t\ge t_0$, the states and input remain bounded, and
\begin{equation}\label{eq:regulation-objective}
\limsup_{t\to\infty}|e(t)|\le M_e,
\end{equation}
where $M_e\ge0$ is uniform over these parameter and initial-state sets.
The regulation is \emph{exact} when $M_e=0$.
A guaranteed finite bound $M_e>0$ gives \emph{approximate regulation}.
The regulation is \emph{practical} if, for every prescribed accuracy $\rho>0$, a controller can be chosen so that $M_e\le\rho$.

We impose the following structural conditions.
Here $\Phi_A(t,s;\mu)$ denotes the state transition matrix generated by $A(\cdot,\mu)$, with $\Phi_\eta(\cdot,\cdot;\mu)=\Phi_{A_\eta}(\cdot,\cdot;\mu)$.
\begin{assumption}[Standing conditions]\label{ass:standing}
The following conditions hold uniformly for $\mu\in\mathcal P$ and $t\ge s\ge t_0\ge0$.
\begin{enumerate}[(i)]
\item The plant has a known, fixed relative degree $r\in\{1,\ldots,n\}$ and can be transformed into the \mbox{Byrnes--Isidori} (BI) form
\begin{equation}\label{eq:bi-form}
\begin{aligned}
\dot\xi_i&=\xi_{i+1}+p_i(t,\mu)w,\quad 1\le i<r,\\
\dot\xi_r&=\alpha(t,\mu)\xi+\gamma(t,\mu)z_p\\
&\quad+b(t,\mu)u+p_r(t,\mu)w,\\
\dot z_p&=A_\eta(t,\mu)z_p+\beta(t,\mu)\xi_1+P_z(t,\mu)w,\\
e&=\xi_1+Q(t,\mu)w,
\end{aligned}
\end{equation}
by a Lyapunov transformation $\operatorname{col}(\xi,z_p)=T(t,\mu)x$, where $\xi=\operatorname{col}(\xi_1,\ldots,\xi_r)$ and $z_p\in\mathbb R^{n-r}$.
The matrices $T(t,\mu)$, $T^{-1}(t,\mu)$, and $\dot T(t,\mu)$ are uniformly bounded.
\item For some constant $M_S>0$, the exosystem satisfies $\|\Phi_S(t,s;\mu)\|\le M_S$.
\item The zero dynamics are uniformly exponentially stable: there exist constants $M_\eta,\lambda_\eta>0$ such that $\|\Phi_\eta(t,s;\mu)\|\le M_\eta e^{-\lambda_\eta(t-s)}$.
\item A known sign $\sigma_b\in\{-1,1\}$ and constants $b_{\min},M_b>0$ satisfy $0<b_{\min}\le\sigma_b b(t,\mu)=|b(t,\mu)|\le M_b$.
\end{enumerate}
\end{assumption}

Norms are Euclidean vector norms and their induced matrix norms unless otherwise specified.
For functions on $[0,\infty)$, write
\begin{equation}\label{eq:cb-norm}
\|f\|_{C_b^s}=\max_{0\le j\le s}\sup_{t\ge0}\|f^{(j)}(t)\|.
\end{equation}
For parameterized functions, a uniform bound also takes the supremum over the indicated parameter set.
The notation $\|\cdot\|_\infty$ for a trajectory refers to its forward interval $[t_0,\infty)$.
Upper bounds are denoted by $M$ with subscripts identifying the corresponding quantities.
Function arguments are omitted in inline expressions and intermediate calculations when the dependence is clear from the context.
For a differentiable matrix $R$, write $\mathcal L_A R=\dot R+RA$.

\subsection{The Ideal-Input Generator}\label{sec:ideal-generator}
The BI coefficients determine an ideal input that generates a zero-error trajectory.
Define
\begin{equation}\label{eq:ideal-rows}
\begin{aligned}
X_1(t,\mu)&=-Q(t,\mu),\\
X_{i+1}(t,\mu)&=(\mathcal L_S X_i)(t,\mu)-p_i(t,\mu),\quad 1\le i<r,\\
X(t,\mu)&=\operatorname{col}\bigl(X_1(t,\mu),\ldots,X_r(t,\mu)\bigr),
\end{aligned}
\end{equation}
and define
\begin{equation}\label{eq:generator-rows}
\begin{aligned}
D_g(t,\mu)&=P_z(t,\mu)-\beta(t,\mu)Q(t,\mu),\\
r_g(t,\mu)&=(\mathcal L_S X_r)(t,\mu)\\
&\quad-\alpha(t,\mu)X(t,\mu)-p_r(t,\mu).
\end{aligned}
\end{equation}

\begin{proposition}[Causal zero-error trajectory]\label{prop:ideal-generator}
Under Assumption~\ref{ass:standing}, for each $\mu\in\mathcal P$, exosystem solution $w(t)$, and initial value $z_{g0}$ in a prescribed bounded set, let
\begin{equation}\label{eq:ideal-trajectory}
\begin{aligned}
\dot z_g(t)&=A_\eta(t,\mu)z_g(t)+D_g(t,\mu)w(t),\\
z_g(t_0)&=z_{g0},\qquad \xi^\star(t)=X(t,\mu)w(t),\\
u^\star(t)&=\frac{r_g(t,\mu)w(t)-\gamma(t,\mu)z_g(t)}{b(t,\mu)}.
\end{aligned}
\end{equation}
Then $(\xi^\star(t),z_g(t),u^\star(t))$ is a bounded trajectory of~\eqref{eq:bi-form} with $e(t)\equiv0$.
The trajectory and its time derivatives needed below are uniformly bounded for $\mu\in\mathcal P$ and $w(t_0),z_{g0}$ in their prescribed bounded sets.
The choice of $z_{g0}$ may depend on $\mu,t_0,w(t_0)$ and need not be known to the controller.
With $x_g(t)=\operatorname{col}(w(t),z_g(t))$, the ideal input is equivalently generated by
\begin{equation}\label{eq:original-generator}
\begin{gathered}
\dot x_g(t)=A_g(t,\mu)x_g(t),\\
v(t)=\ell(t,\mu)x_g(t)=b(t,\mu)u^\star(t),
\end{gathered}
\end{equation}
where
\begin{equation}\label{eq:generator-matrices}
\begin{gathered}
A_g(t,\mu)=\begin{bmatrix}S(t,\mu)&0\\D_g(t,\mu)&A_\eta(t,\mu)\end{bmatrix},\\
\ell(t,\mu)=\begin{bmatrix}r_g(t,\mu)&-\gamma(t,\mu)\end{bmatrix}.
\end{gathered}
\end{equation}
\end{proposition}
\begin{proof}
See Supplementary~S1.1.
\end{proof}
A different bounded initial value of $z_g$ changes the ideal trajectory only by an exponentially decaying zero-dynamics transient and its associated input correction.

The generator~\eqref{eq:original-generator}--\eqref{eq:generator-matrices} is constructed from the BI coefficients without first solving a regulator equation.
It specifies the ideal input that an \emph{internal model} must reproduce.
To relate input reproduction to regulation, define the deviations from the zero-error trajectory by $E_i(t)=\xi_i(t)-X_i(t,\mu)w(t)$, $E(t)=\operatorname{col}(E_1(t),\ldots,E_r(t))$, and $z(t)=z_p(t)-z_g(t)$.
Substitution into~\eqref{eq:bi-form} gives
\begin{equation}\label{eq:error-chain}
\begin{aligned}
\dot z(t)&=A_\eta(t,\mu)z(t)+\beta(t,\mu)e(t),\\
E_1(t)&=e(t),\qquad \dot E_i(t)=E_{i+1}(t),\quad i<r,\\
\dot E_r(t)&=\alpha(t,\mu)E(t)+\gamma(t,\mu)z(t)\\
&\quad+b(t,\mu)\bigl(u(t)-u^\star(t)\bigr).
\end{aligned}
\end{equation}
In particular, for $r=1$,
\begin{equation}\label{eq:error-degree-one}
\begin{aligned}
\dot z(t)&=A_\eta(t,\mu)z(t)+\beta(t,\mu)e(t),\\
\dot e(t)&=\alpha(t,\mu)e(t)+\gamma(t,\mu)z(t)\\
&\quad+b(t,\mu)\bigl(u(t)-u^\star(t)\bigr).
\end{aligned}
\end{equation}

\section{Parameterized Nonmonic Internal Models and Robust Regulation}
\label{sec:nonmonic}
Starting from a \emph{nonmonic differential relation}, Section~\ref{sec:filter-realization} constructs a stable-filter representation and a parameterized readout.
Section~\ref{sec:robust-regulation} uses this realization in an error-feedback controller and establishes robust regulation for a fixed realization and practical regulation for a family of realizations, without persistent excitation.
The relative-degree-one results are followed by an extension to higher relative degree.

\begin{definition}[Nonmonic differential relation]
\label{def:nonmonic}
The ideal-input family admits a uniformly nondegenerate nonmonic differential relation of order $m\ge1$ if known coefficient functions $a_j(t,\mu)$ satisfy
\begin{equation}\label{eq:nonmonic-relation}
\sum_{j=0}^{m}a_j(t,\mu)u^{\star(j)}(t)=\nu_{\mathrm{tr}}(t),
\; a_m(t,\mu)\ge a_{\min}>0,
\end{equation}
for all admissible parameters and ideal trajectories with initial states in the prescribed bounded sets.
The same order and coefficient functions apply throughout this family, independently of the generator initial value.
The residual may depend on the trajectory and satisfies $|\nu_{\mathrm{tr}}(t)|\le M_{\mathrm{tr}}e^{-\lambda_{\mathrm{tr}}(t-t_0)}$, where $M_{\mathrm{tr}}\ge0$ and $\lambda_{\mathrm{tr}}>0$ are uniform over these parameter and initial-state sets.
The case $\nu_{\mathrm{tr}}\equiv0$ gives an exact differential relation.
\end{definition}
The coefficient functions are known, while the value of $\mu$ is unknown.
A uniformly nonzero leading coefficient can be made positive by multiplying the relation by that coefficient; a known negative sign can instead be removed by multiplying by $-1$.

The decaying right-hand side in~\eqref{eq:nonmonic-relation} accommodates different ideal-generator initial values.
For the same parameter and exosystem trajectory, a bounded change of initial value changes the ideal input and its required derivatives by exponentially decaying terms, as proved in Supplementary~S1.1.
Thus, the same relation coefficients remain valid with an additional exponentially decaying residual.

Retaining a nonmonic relation can preserve a finite linear separation of the time and parameter dependence that is lost upon division by the leading coefficient to obtain the monic form.
We use the following definition.

\begin{definition}[Finite separability]\label{def:finite-separability}
A scalar-, vector-, or matrix-valued function $f(t,\mu)$ is \emph{finitely separable} if, for some finite $N$, it admits the representation
\begin{equation*}
f(t,\mu)=f_0(t)+\sum_{i=1}^{N}\theta_i(\mu)f_i(t),
\end{equation*}
where the functions $f_i(t)$ are known and have the same dimensions as $f$, and $\theta:\mathcal P\to\mathbb R^N$ is a known parameter map.
\end{definition}
Affine dependence of $f(t,\mu)$ on $\mu$ is a special case, with $\theta(\mu)=\mu$.
For a fixed unknown $\mu$, write $\theta^\star=\theta(\mu)$ and denote its estimate by $\hat\theta$.
The known time functions and unknown constant weights in this representation enter the parameterized readout in Section~\ref{sec:filter-realization}.
Collecting the parameter functions of finitely many separable coefficients into one vector gives a common representation.
In particular, the coefficients of~\eqref{eq:nonmonic-relation} can then be written as
\begin{equation*}
a_j(t,\mu)=a_{j0}(t)+\sum_{i=1}^{N}\theta_i(\mu)a_{ji}(t),
\qquad 0\le j\le m.
\end{equation*}
Multiplying~\eqref{eq:nonmonic-relation} by a function of $(t,\mu)$ changes both its coefficients and its residual.
The scaled coefficients must retain a positive leading-coefficient bound and finite separability, and the scaled residual must retain a uniform exponential bound.
The following example shows why retaining a nonmonic relation can be essential for finite separability.

\begin{example}\label{ex:nonmonic-separation}
Consider $\dot w=0$, $\dot y=-y+(2+\mu+\sin t)u$, and $e=y-w$, with $|\mu|\le\tfrac12$.
The ideal input $u^\star=w/(2+\mu+\sin t)$ satisfies
\begin{equation*}
(2+\mu+\sin t)\dot u^\star+\cos t\,u^\star=0.
\end{equation*}
The leading coefficient is bounded below by $1/2$, and both coefficients are affine in $\mu$.
Dividing by $(2+\mu+\sin t)$ gives the monic equation $\dot u^\star=-[\cos t/(2+\mu+\sin t)]u^\star$, whose coefficient is not finitely separable.
Moreover, no monic relation of any finite order with finitely separable coefficients generates this entire family, even after a finite constant reparameterization and with a residual tending to zero for each parameter.
The proof is given in Supplementary~S2.
\end{example}

\begin{remark}[Monic and nonmonic differential relations]\label{rm:monic-comparison}
At the level of scalar differential relations, the present nonmonic condition relaxes the exact monic requirement, namely $a_m=1$ with zero right-hand side, used by Zhang and Serrani~\cite[Lemma~3.2 and Assumption~6]{zhang2009tac} and Yang and Huang~\cite[Assumptions~3.2--3.3]{yang2010framework} for their respective ideal inputs.
In both works, finite separability is imposed on the observer-form coefficients. 
Conversion between derivative and observer forms preserves this property.
Thus, at the level of coefficient structure, these assumptions require the ideal input to satisfy a monic relation with finitely separable coefficients.
The nonmonic condition admits a strictly larger class of relations, even when decaying residuals are allowed in the monic case, as shown by Example~\ref{ex:nonmonic-separation}.
\end{remark}

\begin{remark}\label{rm:linear-generation}
Finite separability of the nonmonic relation coefficients does not require finite separation of the ideal-input family, even modulo vanishing terms.
For the known exosystem in~\cite[Theorem~4.1]{zha2026robust}, fixed finite-dimensional linear zero-input generation requires $u^\star=\psi^T\alpha^\star+r_{\mathrm{lin}}$, with known $\psi$, constant $\alpha^\star$ incorporating the exosystem initial state, and $r_{\mathrm{lin}}(t)\to0$.
Example~\ref{ex:nonmonic-separation} admits no such representation, by the argument in~\cite[Example~1]{zha2026robust}, so its ideal-input family cannot be reproduced asymptotically by any such internal model.
Nevertheless, its finitely separable nonmonic relation yields exact regulation by Theorem~\ref{thm:fixed-model-regulation}.
Conversely, exactly finitely separable ideal inputs admit the direct adaptive design in Remark~\ref{rm:direct-input-approximation}.
For $r=1$, the framework also retains the finite-separation-based regulation design in~\cite[Corollary~4.1]{zha2026robust}, with its stabilizability conditions and allowing general vanishing remainders; see Supplementary~S3.3.
It therefore strictly enlarges the corresponding class of exactly regulatable systems.
\end{remark}

\subsection{Stable Filtering and Parameterized Readout}\label{sec:filter-realization}
We first construct a stable filter with a prescribed, parameter-independent Hurwitz matrix to represent $a_m(t,\mu)u^\star(t)$ up to an exponentially decaying transient, without differentiating $u^\star$.
When the relation coefficients are finitely separable, this gives the parallel realization~\eqref{eq:parallel-filter}.
We then formulate conditions for an internal-model realization and bound its model residual in terms of coefficient-approximation errors.

\subsubsection{Stable Filtering}
We begin with a stable-filter representation of~\eqref{eq:nonmonic-relation}.
Let $d(t,\mu)=a_m(t,\mu)$, let $\mathscr D=d/dt$ denote differentiation in time, and define
\begin{equation}\label{eq:rearranged-coefficients}
\mathfrak c_j(t,\mu)=\sum_{l=j}^{m}(-1)^{l-j}\binom lj
a_l^{(l-j)}(t,\mu),\quad 0\le j\le m.
\end{equation}
Choose a monic Hurwitz polynomial and the corresponding coefficients
\begin{equation}\label{eq:filter-polynomial}
\begin{aligned}
\mathsf P(s)&=s^m+\sum_{j=0}^{m-1}f_js^j,\\
h_j(t,\mu)&=f_jd(t,\mu)-\mathfrak c_j(t,\mu)\quad(j<m).
\end{aligned}
\end{equation}
Products of time-varying coefficients and differential operators are ordered as written; coefficients are not commuted through a filter.

\begin{proposition}[Forward filter realization]\label{prop:forward-filter}
Relation~\eqref{eq:nonmonic-relation} implies
\begin{equation}\label{eq:filtered-identity}
\begin{aligned}
\mathsf P(\mathscr D)[d(t,\mu)u^\star(t)]
&=\nu_{\mathrm{tr}}(t)+\sum_{j=0}^{m-1}\mathscr D^j[h_j(t,\mu)u^\star(t)].\\
\end{aligned}
\end{equation}
Let $F_c$ be the companion matrix with ones on the superdiagonal and last row $(-f_0,\ldots,-f_{m-1})$.
Set $F_0=F_c^T$, $C_0=\mathbf e_m^T$, and $B_j=\mathbf e_{j+1}$, where $\mathbf e_i$ is the $i$th standard basis vector of $\mathbb R^m$.
Then
\begin{equation}\label{eq:filter-transfer}
C_0(sI-F_0)^{-1}B_j=\frac{s^j}{\mathsf P(s)}=:Q_j(s).
\end{equation}
There exists a uniformly bounded analysis initial value such that
\begin{equation}\label{eq:ideal-filter}
\begin{aligned}
\dot\eta^\star(t)&=F_0\eta^\star(t)+\sum_{j=0}^{m-1}B_jh_j(t,\mu)u^\star(t),\\
C_0\eta^\star(t)&=d(t,\mu)u^\star(t)-\tau_{\mathrm{tr}}(t)
\end{aligned}
\end{equation}
holds on $[t_0,\infty)$, where $\tau_{\mathrm{tr}}(t)=\int_{t_0}^t q_{\mathrm{tr}}(t-s)\nu_{\mathrm{tr}}(s)\,ds$ and $q_{\mathrm{tr}}$ is the impulse response of $1/\mathsf P(s)$.
This transient decays exponentially and belongs to $L^2\cap L^\infty$, with uniform bounds over the prescribed parameter and initial-state sets.
\end{proposition}
\begin{proof}
The product rule yields~\eqref{eq:filtered-identity}.
The stated realization follows by absorbing the homogeneous filter-output mismatch into its ideal initial state.
The construction and uniform bounds are given in Supplementary~S1.2.
\end{proof}
\begin{remark}\label{rm:filter-realization}
Proposition~\ref{prop:forward-filter} converts a nonmonic relation into a stable-filter input--output identity with an exponentially decaying residual.
A compatible initial value absorbs the homogeneous mismatch into the ideal filter state, leaving only the zero-state response to $\nu_{\mathrm{tr}}$ in the readout residual $\tau_{\mathrm{tr}}$.
Any other bounded ideal initial value adds an exponentially decaying homogeneous term.
Moreover, if the relation coefficients are finitely separable, the construction gives the parallel realization below with known, parameter-independent filter dynamics and unknown constant weights in the readout.
\end{remark}

For this realization, set $d(t,\theta)=a_{m0}(t)+\sum_{i=1}^N\theta_i a_{mi}(t)$, so that $d(t,\theta^\star)=a_m(t,\mu)$.
Writing $h_j(t,\mu)=h_{j0}(t)+\sum_{i=1}^N\theta_i(\mu)h_{ji}(t)$ gives
\begin{equation}\label{eq:parallel-filter}
\begin{aligned}
\dot\zeta_i(t)&=F_0\zeta_i(t)+G_i(t)u(t),\quad 0\le i\le N,\\
G_i(t)&=\sum_{j=0}^{m-1}B_jh_{ji}(t),\\
h(t,\theta)\zeta(t)&=C_0\zeta_0(t)+\sum_{i=1}^N\theta_iC_0\zeta_i(t),
\end{aligned}
\end{equation}
where $\zeta(t)=\operatorname{col}(\zeta_0(t),\ldots,\zeta_N(t))$.
The known filter matrices depend only on time.
With compatible ideal initial conditions, its readout satisfies
\begin{equation}\label{eq:exact-readout}
d(t,\theta^\star)u^\star(t)=h(t,\theta^\star)\zeta^\star(t)+\tau_{\mathrm{tr}}(t).
\end{equation}
The parameterized readout uses the ratio $h(t,\hat\theta(t))\zeta(t)/d(t,\hat\theta(t))$, while its numerator and denominator retain their affine dependence on the adjustable parameter.
The error-feedback update law is given in Section~\ref{sec:robust-regulation}.

\subsubsection{Parameterized Readout}
We use the finite-dimensional stable filter $\dot\zeta_K(t)=F_K(t)\zeta_K(t)+G_K(t)u(t)$, whose known matrices are independent of the adjustable parameter, and the parameterized readout $h_K(t,\theta)\zeta_K(t)/d_K(t,\theta)$ as an internal-model realization.
Together with the parameter update law introduced below, they form the adaptive internal model.
The parallel realization~\eqref{eq:parallel-filter} is a special case of this form.
Fix one realization index $K$ and write
\begin{equation}\label{eq:affine-model}
\begin{aligned}
d_K(t,\theta)&=d_{0,K}(t)+\sum_{i=1}^{N_K}\theta_i d_{i,K}(t),\\
h_K(t,\theta)&=h_{0,K}(t)+\sum_{i=1}^{N_K}\theta_i h_{i,K}(t),
\end{aligned}
\end{equation}
where $h_K$ is a row vector acting on the filter state.

\begin{assumption}[Filter and readout conditions]\label{ass:finite-model}
For the fixed $K$, the following properties hold.
\begin{enumerate}[(i)]
\item The known matrix $F_K(t)$ is uniformly exponentially stable.
The affine coefficient maps in~\eqref{eq:affine-model} extend to a neighborhood of $\Theta_K$ with uniformly bounded first derivatives there.
\item A known, implementable, nonempty compact convex set $\Theta_K\subset\mathbb R^{N_K}$ satisfies $d_K(t,\theta)\ge d_{\min}>0$ for all $t\ge0$ and $\theta\in\Theta_K$.
\item For every allowed parameter and ideal trajectory, there is a constant comparison parameter $\theta_K^\star\in\Theta_K$ and a bounded ideal filter trajectory with bounded initial value such that
\begin{equation*}
\begin{aligned}
\dot\zeta_K^\star(t)&=F_K(t)\zeta_K^\star(t)+G_K(t)u^\star(t),\\
\Delta_{\mathrm p,K}(t)+\tau_K(t)&=d_K(t,\theta_K^\star)u^\star(t)\\
&\quad-h_K(t,\theta_K^\star)\zeta_K^\star(t),\\
\|\Delta_{\mathrm p,K}\|_\infty&\le\varepsilon_K,\qquad
\tau_K\in L^2\cap L^\infty.
\end{aligned}
\end{equation*}
\end{enumerate}
The pointwise bounds are uniform in time and in the prescribed physical parameter and initial-state sets, and the $L^2$ norm of $\tau_K$ is uniformly bounded over these sets.
These bounds, including the bound on the comparison initial state, may depend on the fixed realization $K$.
No approximation family is required.
\end{assumption}

\begin{remark}\label{rm:model-residuals}
Assumption~\ref{ass:finite-model} concerns a fixed finite-dimensional filter and a readout with $N_K$ adjustable parameters in $\Theta_K$.
The actual filter may be initialized at zero, and parameter estimates are initialized in $\Theta_K$.
For each admissible ideal trajectory, a constant comparison parameter must give the stated residual bound.
The comparison parameter $\theta_K^\star$ need not equal the physical parameter or its exact lifting.
Here $\Delta_{\mathrm p,K}$ describes persistent model mismatch, while $\tau_K$ accounts for finite-energy transients and need not decay exponentially.
The fixed-dead-zone results below require the uniform finite-energy bound but not its numerical value.
The additional numerical bounds used for automatic dead-zone adjustment are specified in Supplementary~S5.
Thus the realization conditions allow both zero persistent residual and a bounded persistent residual, as needed for exact and practical regulation, respectively.
\end{remark}

\begin{remark}\label{prop:linear-im-inclusion}
The parallel realization~\eqref{eq:parallel-filter} satisfies Assumption~\ref{ass:finite-model} under Proposition~\ref{prop:forward-filter} when the candidate set in item~(ii) contains $\theta(\mathcal P)$, with $\Delta_{\mathrm p,K}=0$ and $\tau_K=\tau_{\mathrm{tr}}$.
The assumption also covers a known LTV canonical realization $(F_{\mathrm{im}},G_{\mathrm{im}},H_{\mathrm{im}})$ of the ideal input, as defined in~\cite[Definition~B.1]{zha2026robust}.
For exact reproduction, take $F_K=F_{\mathrm{im}}$, $G_K=G_{\mathrm{im}}$, $h_K=H_{\mathrm{im}}$, $d_K=1$, and $N_K=0$, giving zero readout residual.
The verification is given in Supplementary~S3.1.
\end{remark}

\begin{remark}\label{rm:coefficient-residual}
When exact finite separation is unavailable, a finite coefficient approximation can be used if it satisfies the candidate-set and realization requirements in Assumption~\ref{ass:finite-model}.
Its coefficient error gives a bound on the persistent model residual as follows.
For a coefficient approximation to a fixed relation and filter template, write $d_K(t,\theta_K^\star)$ and $h^{\rm rel}_{j,K}(t,\theta_K^\star)$ for approximations of the scalar coefficients $d(t,\mu)$ and $h_j(t,\mu)$ in~\eqref{eq:filter-polynomial}, evaluated at the comparison parameter.
The scalars $h^{\rm rel}_{j,K}(t,\theta_K^\star)$ are distinct from the readout rows $h_{i,K}(t)$ in~\eqref{eq:affine-model}.
Let $\varepsilon_{\mathrm{coef},K}$ bound all these scalar coefficient errors uniformly.
The common effective ideal initialization in Supplementary~S1.2 gives
\begin{equation}\label{eq:coefficient-to-residual}
\|\Delta_{\mathrm p,K}\|_\infty\le M_{u^\star} M_Q\varepsilon_{\mathrm{coef},K},
\end{equation}
where $M_Q=1+\sum_{j=0}^{m-1}\|q_j\|_{L^1}$, $\|u^\star\|_\infty\le M_{u^\star}$, and $q_j$ is the impulse response of $Q_j$.
One may therefore choose $\varepsilon_K=M_{u^\star}M_Q\varepsilon_{\mathrm{coef},K}$.
This estimate bounds the persistent residual; the filtered relation residual and homogeneous initialization terms belong to $\tau_K$.
The constant $M_Q$ depends on the fixed closure order and filter template, not on the number of parameter terms.
\end{remark}

\subsection{Robust Regulation Without Persistent Excitation}
\label{sec:robust-regulation}
We consider $r=1$ and fix an internal-model realization satisfying Assumption~\ref{ass:finite-model}.
For $r\ge2$, see Section~\ref{sec:higher-relative-degree}.
The controller drives the filter by $u$ and updates its parameters from the measured regulation error $e$.

\subsubsection{Controller Design}
Suppress the fixed index $K$ in the coefficients, retain $\varepsilon_K$ for the persistent-residual bound, and write $\tilde\theta(t)=\hat\theta(t)-\theta^\star$.
For a scalar $s$ and $\delta\ge0$, define the dead-zone function
\begin{equation}\label{eq:dead-zone}
\mathcal D_\delta(s)=\operatorname{sign}(s)(|s|-\delta)_+,
\qquad (a)_+=\max\{a,0\}.
\end{equation}
For a constant symmetric matrix $\Gamma>0$, use the metric projection of a velocity onto the tangent cone $T_\Theta(\theta)$:
\begin{equation}\label{eq:metric-projection}
\operatorname{Proj}_\Theta^\Gamma(\theta,v)
=\mathop{\arg\min}_{w\in T_\Theta(\theta)}
(w-v)^T\Gamma^{-1}(w-v).
\end{equation}
The projection inequalities and the local Carath\'eodory solution framework are given in Supplementary~S1.3.

For a gain $k>0$, a fixed dead zone $\delta\ge0$, and a constant symmetric matrix $\Gamma>0$, consider the controller
\begin{equation}\label{eq:r1-controller}
\begin{aligned}
\widehat H(t)&=\frac{h(t,\hat\theta(t))}{d(t,\hat\theta(t))},\\
u(t)&=\widehat H(t)\zeta(t)-k\sigma_b e(t),\\
\dot\zeta(t)&=F(t)\zeta(t)+G(t)u(t),\\
\phi_i(t)&=d_i(t)u(t)-h_i(t)\zeta(t),\quad 1\le i\le N,\\
\dot{\hat\theta}(t)&=\operatorname{Proj}_\Theta^\Gamma
\bigl(\hat\theta(t),\sigma_b\Gamma\phi(t)\mathcal D_\delta(e(t))\bigr),
\end{aligned}
\end{equation}
where $\phi=\operatorname{col}(\phi_1,\ldots,\phi_N)$, and the dead-zone function and projection are defined in~\eqref{eq:dead-zone} and~\eqref{eq:metric-projection}.
The denominator is positive on $\Theta$.
At each time, the current error and controller states determine $u(t)$, after which $\phi(t)$ and the update rate are evaluated.
The filter may start from any finite state, including zero, and $\hat\theta(t_0)\in\Theta$.

For the comparison parameter and trajectory in Assumption~\ref{ass:finite-model}, write $d^\star(t)=d(t,\theta^\star)$, $h^\star(t)=h(t,\theta^\star)$, $\widehat d(t)=d(t,\hat\theta(t))$, and $\tilde\zeta(t)=\zeta(t)-\zeta^\star(t)$.
The affine parameterization gives the matching identity
\begin{equation}\label{eq:r1-matching}
\begin{aligned}
d^\star(u-u^\star)
={}&h^\star\tilde\zeta-k\sigma_b\widehat d\,e\\
&-\phi^T\tilde\theta-\Delta_{\mathrm p,K}-\tau_K.
\end{aligned}
\end{equation}
In the construction~\eqref{eq:parallel-filter}, retaining the nonmonic leading coefficient preserves the finite separation underlying this affine matching structure.
The term $-\phi^T\tilde\theta$ in~\eqref{eq:r1-matching} is matched by the parameter update in~\eqref{eq:r1-controller}, so that the parameter cross term cancels in the storage estimate below.

\subsubsection{Robust Regulation}
To specify the gain conditions, define the analysis variables
\begin{equation}\label{eq:r1-transformation}
\begin{aligned}
R_f(t,\mu)&=\frac{G(t)}{b(t,\mu)},\\
q_f(t)&=\tilde\zeta(t)-R_f(t,\mu)e(t),\\
\chi(t)&=\operatorname{col}(z(t),q_f(t)),\\
p(t,\mu)&=\frac{d^\star(t)}{\sigma_b b(t,\mu)}.
\end{aligned}
\end{equation}
The transformed state $\chi$ satisfies a stable cascade driven by $e$.
Appendix~\ref{app:fixed-model-regulation} constructs a uniformly positive definite matrix $P_\chi(t,\mu)$ for this cascade, so that
\begin{equation}\label{eq:r1-comparison-norm}
\begin{gathered}
W(t)=\bigl(\chi(t)^TP_\chi(t,\mu)\chi(t)\bigr)^{1/2},\\
D^+W(t)\le-\lambda_\chi W(t)+M_{\chi e}|e(t)|,
\end{gathered}
\end{equation}
where $\lambda_\chi>0$, $M_{\chi e}\ge0$, and $D^+$ denotes the upper right Dini derivative.
Set
\begin{equation*}
\begin{aligned}
A_e(t,\mu)&=p(t,\mu)\alpha(t,\mu)
              +\frac{h^\star(t)G(t)}{\sigma_b b(t,\mu)},\\
B_e(t,\mu)&=\begin{bmatrix}p(t,\mu)\gamma(t,\mu)&\sigma_b h^\star(t)\end{bmatrix},
\end{aligned}
\end{equation*}
and choose uniform bounds
\begin{equation}\label{eq:r1-coefficient-bounds}
\begin{gathered}
0<p_{\min}\le p(t,\mu)\le M_p,\qquad
|\dot p(t,\mu)|\le M_{\dot p},\\
|A_e(t,\mu)|\le M_{A_e},\qquad
|B_e(t,\mu)\chi(t)|\le M_{B_e} W(t).
\end{gathered}
\end{equation}
These constants are obtained from the bounds for the plant and the fixed realization, uniformly over the admissible parameters and comparison trajectories.
They are independent of $k$, $\delta$, and $\Gamma$, but may depend on the fixed realization $K$.
Explicit choices are given in Appendix~\ref{app:fixed-model-regulation}.

\begin{theorem}[Robust regulation for relative degree one]\label{thm:fixed-model-regulation}
Suppose Assumptions~\ref{ass:standing} and~\ref{ass:finite-model} hold with $r=1$.
Fix $\Gamma>0$ and choose $k>0$ such that
\begin{equation}\label{eq:r1-gain-conditions}
\begin{aligned}
c_e&:=kd_{\min}-M_{A_e}-\frac{M_{\dot p}}{2}
              -\frac{(M_{\chi e}+M_{B_e})^2}{2\lambda_\chi}>0,\\
\kappa&:=kd_{\min}-M_{A_e}-\frac{M_{B_e}M_{\chi e}}{\lambda_\chi}>0.
\end{aligned}
\end{equation}
For every $\mu\in\mathcal P$, plant and exosystem initial states in the prescribed sets, finite $\zeta(t_0)$, and $\hat\theta(t_0)\in\Theta$, controller~\eqref{eq:r1-controller} has a unique closed-loop solution on $[t_0,\infty)$ for every fixed finite $\delta\ge0$.
The closed-loop states and input are bounded.
If the dead zone is selected after $k$ so that
\begin{equation}\label{eq:r1-dead-zone-condition}
\kappa\delta\ge\varepsilon_K,
\end{equation}
then
\begin{equation}\label{eq:r1-regulation-bound}
\limsup_{t\to\infty}|e(t)|\le\delta.
\end{equation}
For zero persistent residual, $\Delta_{\mathrm p,K}\equiv0$, and $\delta=0$, one has
\begin{equation}\label{eq:r1-exact-convergence}
\begin{gathered}
e(t)\longrightarrow0,\qquad u(t)-u^\star(t)\longrightarrow0,\\
\widehat H(t)\zeta(t)-u^\star(t)\longrightarrow0.
\end{gathered}
\end{equation}
These conclusions require neither persistent excitation nor convergence of the parameter estimates.
\end{theorem}

\begin{proof}
The matching identity explains the choice of the update sign and the weight $p(t,\mu)$.
With $E_\delta(t)=\mathcal D_\delta(e(t))$ and $X_\delta(t)=(W(t)-M_{\chi e}\delta/\lambda_\chi)_+$, the proof uses
\begin{equation}\label{eq:r1-storage}
V_\delta(t)=\frac{X_\delta(t)^2}{2}
+\frac{p(t,\mu)E_\delta(t)^2}{2}
+\frac{\tilde\theta(t)^T\Gamma^{-1}\tilde\theta(t)}{2}.
\end{equation}
The projection inequality cancels the parameter cross term from~\eqref{eq:r1-matching}, yielding
\begin{equation}\label{eq:r1-dissipation}
\begin{aligned}
\dot V_\delta\le{}&-\frac{\lambda_\chi}{2}X_\delta^2-c_eE_\delta^2\\
&-(\kappa\delta-\varepsilon_K)|E_\delta|+|\tau_K||E_\delta|
\end{aligned}
\end{equation}
almost everywhere.
Under~\eqref{eq:r1-dead-zone-condition}, Young's inequality gives
\begin{equation}\label{eq:r1-transient-dissipation}
\dot V_\delta\le-\frac{\lambda_\chi}{2}X_\delta^2-\frac{c_e}{2}E_\delta^2
+\frac{|\tau_K|^2}{2c_e}.
\end{equation}
Appendix~\ref{app:fixed-model-regulation} proves this estimate and derives boundedness, global continuation, and convergence.
\end{proof}

\begin{remark}\label{rm:fixed-model-accuracy}
For a fixed gain satisfying~\eqref{eq:r1-gain-conditions}, the choice $\delta=\varepsilon_K/\kappa$ gives an error bound proportional to the persistent residual.
The dead zone stops parameter adaptation when $|e|\le\delta$; the condition $\kappa\delta\ge\varepsilon_K$ allows the storage estimate to absorb persistent model mismatch outside this band.
This conclusion concerns regulation and does not require identification of the comparison parameter.
\end{remark}

\begin{remark}[Practical regulation by high-gain feedback]
\label{rm:high-gain-comparison}
Under Assumption~\ref{ass:standing} with $r=1$, practical regulation can also be achieved by increasing $k$ for a fixed realization satisfying Assumption~\ref{ass:finite-model}(i)--(ii), even when the internal model does not reproduce $u^\star$.
The filter and readout bounds are held fixed as $k$ increases.
Omitting the feedforward term gives the special case $u=-k\sigma_b e$.
The estimates and proof are given in Supplementary~S4.
\end{remark}

\subsubsection{Practical Regulation}
Theorem~\ref{thm:fixed-model-regulation} concerns a fixed realization.
For $r=1$, we now approximate the same ideal-input family by increasingly accurate finite realizations while keeping the stabilization gain fixed.
Under the common bounds stated below, one gain $k$ satisfying~\eqref{eq:r1-gain-conditions} gives a common $\kappa>0$, independent of $K$.
As the persistent-residual bound $\varepsilon_K\to0$, the choice $\delta_K=2\varepsilon_K/\kappa$ then makes the guaranteed ultimate error arbitrarily small.
Choose constant symmetric matrices $\Gamma_K>0$ by a prescribed rule whose scalar design parameters are held fixed.

Section~\ref{sec:approximation-construction} gives a general coefficient-approximation framework and verifies the following conditions through common bounds on the comparison coefficient combinations, using Taylor expansions as the main construction.

\begin{assumption}\label{ass:model-family}
For some $K_0$, the following conditions hold for all $K\ge K_0$.
\begin{enumerate}[(i)]
\item Each realization satisfies Assumptions~\ref{ass:standing} and~\ref{ass:finite-model} with $r=1$.
In particular, it has a constant comparison parameter in its known compact convex set $\Theta_K$, and
\begin{equation*}
\begin{gathered}
\|\Delta_{\mathrm p,K}\|_\infty\le\varepsilon_K,\qquad
\varepsilon_K\longrightarrow0,\\
\tau_K\in L^2\cap L^\infty.
\end{gathered}
\end{equation*}
The transient and comparison initial-state bounds are uniform over the prescribed data for each fixed $K$, but may depend on $K$.
\item Retain $p_K,A_{e,K},B_{e,K},\chi_K$ from~\eqref{eq:r1-transformation}--\eqref{eq:r1-coefficient-bounds}.
There are constants $d_{\min},\lambda_\chi>0$ and $M_{\dot p},M_{A_e},M_{B_e},M_{\chi e}\ge0$, independent of $K$, such that
\begin{equation}\label{eq:family-comparison-bounds}
\begin{gathered}
d_K(t,\theta)\ge d_{\min}\quad(t\ge0,\ \theta\in\Theta_K),\\
|\dot p_K|\le M_{\dot p},\qquad |A_{e,K}|\le M_{A_e},\\
|B_{e,K}\chi_K|\le M_{B_e}W_K,\\
D^+W_K\le-\lambda_\chi W_K+M_{\chi e}|e|.
\end{gathered}
\end{equation}
Here $W_K\ge0$ is an analysis quantity that is locally absolutely continuous along solutions of the stable analysis system, with finite initial values.
These estimates hold uniformly in time and over the prescribed physical parameters and comparison trajectories, for arbitrary analysis states and locally bounded inputs $e$.
They do not assume boundedness of the closed-loop trajectory.
The quantity $W_K$ may use the full transformed state or effective combinations formed with the constant comparison parameter, provided that the actual coupling $B_{e,K}\chi_K$ satisfies the stated bound.
\end{enumerate}
\end{assumption}
For every $K$, $p_K\ge p_{\min}=d_{\min}/M_b$; an upper bound $M_{p,K}$ need only be finite for each $K$.

\begin{theorem}[Practical regulation]\label{thm:model-family-regulation}
Suppose Assumption~\ref{ass:model-family} holds.
Choose one $k$ satisfying~\eqref{eq:r1-gain-conditions} with the common bounds~\eqref{eq:family-comparison-bounds}.
Choose fixed dead zones $\delta_K\ge0$ such that
\begin{equation}\label{eq:family-dead-zone}
\kappa\delta_K\ge\varepsilon_K,\qquad \delta_K\longrightarrow0.
\end{equation}
For every fixed $K\ge K_0$, every admissible plant and exosystem initial state, arbitrary finite actual filter initial states, and $\hat\theta_K(t_0)\in\Theta_K$, controller~\eqref{eq:r1-controller} has a unique solution on $[t_0,\infty)$.
All closed-loop states and the input are bounded, and
\begin{equation}\label{eq:family-regulation}
\begin{gathered}
\limsup_{t\to\infty}|e_K(t)|\le\delta_K,\\
\lim_{K\to\infty}\limsup_{t\to\infty}|e_K(t)|=0.
\end{gathered}
\end{equation}
In particular, the conservative choice $\delta_K=2\varepsilon_K/\kappa$ achieves any prescribed accuracy $\rho>0$ by choosing $K$ offline so that $\varepsilon_K\le\kappa\rho/2$.
\end{theorem}
\begin{proof}
The common comparison bounds give one choice of the stabilization constants.
The storage estimates absorb the initial storage and transient energy for each fixed $K$; their sizes need not be uniform in $K$.
Appendix~\ref{app:model-family-regulation} proves boundedness and the first inequality in~\eqref{eq:family-regulation}, including when $W_K$ uses only effective states.
The iterated limit then follows from $\delta_K\to0$.
\end{proof}
\begin{remark}\label{rm:family-fixed-design}
Theorem~\ref{thm:model-family-regulation} permits the stabilization gain $k$ to be fixed before selecting a more accurate finite realization, which remains fixed during each run.
The matrices $\Gamma_K$ follow the prescribed normalization rule and may have different sizes.
The error bound holds for every admissible parameter and initial condition, while state peaks, transient energies, and entry times may depend on $K$ and the trajectory.
\end{remark}

\begin{remark}\label{rm:family-constructions}
For families obtained by approximating a nonmonic relation, keep its order and relation-filter template fixed while increasing the accuracy of the finite parameter representation.
Remark~\ref{rm:coefficient-residual} supplies~\eqref{eq:coefficient-to-residual}.
The coefficient error bounds the denominator and the reordered scalar coefficients, including the derivative errors required in their formation.
Initialization and filtered relation transients remain in $\tau_K$.
An independently specified realization family need not admit a common scalar differential relation.
Section~\ref{sec:closure} gives sufficient conditions for such relations from the original system, and Section~\ref{sec:approximation-construction} verifies the family conditions through coefficient approximation, with a weighted Taylor construction and the convex interpolation alternative in Remark~\ref{rm:convex-interpolation}.
\end{remark}

\begin{remark}[Automatic dead-zone adjustment]\label{rm:auto-dead-zone}
For a fixed realization, the dead-zone width can be increased online without using the numerical value of $\varepsilon_K$.
Under the computable initial-state and transient-energy bounds specified in Supplementary~S5, only finitely many increases occur, and the closed-loop states and input remain bounded.
The initial dead-zone width $\delta_0>0$ may be chosen arbitrarily small, and the ultimate error bound is $\max\{\delta_0,2\varepsilon_K/\kappa\}$, while the realization and stabilization gain remain fixed.
\end{remark}

\subsubsection{Regulation for Higher Relative Degree}\label{sec:higher-relative-degree}\label{rm:higher-relative-degree}
For a known fixed $r\ge2$, the full construction, additional regularity conditions, and proofs are given in Supplementary~S6 and~S7.
We summarize the main steps and regulation results here.
Set $n_r=r-1$, fix $\lambda>0$, and introduce the stable input filter
\begin{equation}\label{eq:hr-input-filter}
\begin{aligned}
\dot\rho_i(t)&=-\lambda\rho_i(t)+\rho_{i+1}(t),\quad 1\le i<n_r,\\
\dot\rho_{n_r}(t)&=-\lambda\rho_{n_r}(t)+u(t).
\end{aligned}
\end{equation}
Driving the same filter by $u^\star$ from a bounded ideal initial value gives a target $\rho_1^\star$ satisfying $\Lambda(\mathscr D)\rho_1^\star=u^\star$, where $\Lambda(s)=(s+\lambda)^{n_r}$.
The right-composed relation $\mathcal A(t,\mu,\mathscr D)\Lambda(\mathscr D)\rho_1^\star=\nu_{\mathrm{tr}}$, with $\mathcal A=\sum_{j=0}^m a_j(t,\mu)\mathscr D^j$, retains the nonmonic form and finite separability, with the same leading coefficient and no differentiation of the residual.
The realization conditions are imposed on $\rho_1^\star$.
The input filter permits the error dynamics to be reduced to a first-order error equation coupled to a stable analysis system and driven by $(\rho_1-\rho_1^\star)$, as in~\textup{(S30)}.
The actual internal-model filter and its readout are
\begin{equation*}
\dot\zeta=F(t)\zeta+G(t)\rho_1,\qquad
a_c=\frac{h(t,\hat\theta)\zeta}{d(t,\hat\theta)}.
\end{equation*}
Starting from $v_1=e+(\sigma_b/k)(\rho_1-a_c)$, successive coordinates are formed by differentiating along the known filter dynamics with $e$ and $\hat\theta$ held fixed and adding the prescribed damping and adjacent-coordinate terms.
At each step, the damping in~\textup{(S40)} is chosen before the next coordinate in~\textup{(S33)} is formed.
The projected update uses a radial dead zone based on $(e,v_1,\ldots,v_{n_r})$ and a regressor formed with $\rho_1$.
It is evaluated before the actual input in~\textup{(S34)}, which cancels the last coordinate's parameter-update term.
The construction uses only time, the regulation error, and controller states, without differentiating the projected update.

Under the hypotheses of Supplementary Theorem~S1, all closed-loop states and the actual input are bounded for every fixed finite $\delta\ge0$.
With the prescribed $\kappa_r=\kappa_*/\sqrt r>0$, the choice $\kappa_r\delta\ge\varepsilon_{r,K}$ gives $\limsup_{t\to\infty}|e(t)|\le\delta$; zero persistent residual with zero dead zone gives exact regulation.
For a realization family satisfying the common comparison bounds and uniform finite-order realization conditions of Supplementary Theorem~S2, keep $\lambda,k,a_*,\kappa_*$ and the damping rule fixed.
Then $\varepsilon_{r,K}\to0$ and $\delta_K=2\varepsilon_{r,K}/\kappa_r$ give $\lim_{K\to\infty}\limsup_{t\to\infty}|e_K(t)|=0$, achieving practical regulation with a finite realization selected before each run.

\section{Existence and Construction of Nonmonic Differential Relations}\label{sec:closure}\label{sec:closure-ltv}\label{sec:closure-periodic}
This section constructs a uniformly nondegenerate differential relation in Definition~\ref{def:nonmonic} from the BI and exosystem coefficients through~\eqref{eq:original-generator}--\eqref{eq:generator-matrices}.
Lemma~\ref{lem:gram-relation} gives a finite-order criterion and an explicit coefficient construction.
Furthermore, Propositions~\ref{prop:ltv-relation} and~\ref{prop:periodic-relation} give sufficient conditions for general LTV and periodic systems, respectively.

Let $n_g=\dim x_g$ and retain the output $v=\ell x_g=bu^\star$ of~\eqref{eq:original-generator}.
Working with $v$ avoids the reciprocal of the uncertain scalar gain $b$ in the output row and its derivative recursion; the relation for $u^\star$ follows by substituting $v=bu^\star$.
Define the output derivative rows, for $j\ge0$, and their finite prefixes by
\begin{equation}\label{eq:generator-derivative-rows}
\begin{aligned}
\ell_j(t,\mu)&=(\mathcal L_{A_g}^{\,j}\ell)(t,\mu),\\
J_m(t,\mu)&=\operatorname{col}(\ell_0(t,\mu),\ldots,\ell_{m-1}(t,\mu)).
\end{aligned}
\end{equation}
Along every generator trajectory, $v^{(j)}=\ell_jx_g$.

\begin{lemma}[Finite-order Gram construction]\label{lem:gram-relation}
Under Assumption~\ref{ass:standing}(iv), suppose $\ell,b\in C_b^m$ and $A_g\in C_b^{m-1}$ uniformly over $\mathcal P$, for a fixed finite $m\ge1$.
Suppose further that a fixed integer $r_o\ge1$ and a constant $\gamma_J>0$ satisfy
\begin{equation}\label{eq:finite-derivative-certificate}
\begin{gathered}
\operatorname{rank}J_m(t,\mu)=\operatorname{rank}J_{m+1}(t,\mu)=r_o,\\
\lambda_{\min}^{+}\bigl(J_m(t,\mu)^TJ_m(t,\mu)\bigr)\ge\gamma_J
\end{gathered}
\end{equation}
for every $t\ge0$ and $\mu\in\mathcal P$, where $\lambda_{\min}^{+}$ denotes the smallest positive eigenvalue.
Set $W_m=J_m^TJ_m$ and let $e_j(W_m)$ be the sum of all principal minors of order $j$, with $e_0=1$.
Define
\begin{equation}\label{eq:gram-polynomial}
\begin{aligned}
\delta_J(t,\mu)&=e_{r_o}(W_m(t,\mu)),\\
B_{r_o}(W_m)&=\sum_{j=0}^{r_o-1}(-1)^j e_{r_o-1-j}(W_m)W_m^j,\\
[\beta_0,\ldots,\beta_{m-1}]&=\ell_m B_{r_o}(W_m)J_m^T.
\end{aligned}
\end{equation}
Then every generator output satisfies
\begin{equation}\label{eq:gram-output-relation}
\delta_Jv^{(m)}-\sum_{j=0}^{m-1}\beta_jv^{(j)}=0,
\qquad \delta_J\ge\gamma_J^{r_o}.
\end{equation}
Substitution of $v=bu^\star$ and multiplication of the entire relation by $b$ give $\sum_{j=0}^m a_j u^{\star(j)}=0$, where, for $0\le j\le m$,
\begin{equation}\label{eq:gram-input-coefficients}
a_j(t,\mu)=b\biggl[\delta_J\binom mj b^{(m-j)}-\sum_{l=j}^{m-1}\beta_l\binom lj b^{(l-j)}\biggr].
\end{equation}
The sum is empty for $j=m$, and Assumption~\ref{ass:standing}(iv) gives
\begin{equation}\label{eq:gram-leading-bound}
a_m(t,\mu)=\delta_J(t,\mu)b(t,\mu)^2
\ge\gamma_J^{r_o}b_{\min}^2>0.
\end{equation}
\end{lemma}
\begin{proof}
\normalfont
The nonzero eigenvalues $\lambda_1,\ldots,\lambda_{r_o}$ of $W_m$ satisfy the polynomial
$(s^{r_o}-e_1s^{r_o-1}+\cdots+(-1)^{r_o}e_{r_o})$.
Thus $\delta_J=\prod_{i=1}^{r_o}\lambda_i\ge\gamma_J^{r_o}$ and spectral decomposition gives $W_mB_{r_o}(W_m)=\delta_J P_{\operatorname{im}W_m}$, where $P_{\operatorname{im}W_m}$ is the orthogonal projection onto the image of $W_m$.
The two factors commute because $B_{r_o}(W_m)$ is a polynomial in $W_m$.
The adjacent-rank equality places $\ell_m$ in the row space of $J_m$, so $\ell_mB_{r_o}(W_m)J_m^TJ_m=\delta_J\ell_m$.
Multiplying by any $x_g$ proves~\eqref{eq:gram-output-relation}.
Leibniz's formula applied to $v=bu^\star$, followed by multiplication by $b$, gives~\eqref{eq:gram-input-coefficients} and~\eqref{eq:gram-leading-bound}.
\end{proof}

\begin{remark}\label{rm:relation-initialization}
This is a row identity, so the exact relation has $\nu_{\mathrm{tr}}=0$ for every bounded ideal initial-state family allowed in Proposition~\ref{prop:ideal-generator}, including parameter-dependent choices of $z_g(t_0)$.
Subsequent coefficient approximation can introduce a persistent residual, while a different ideal filter initialization can introduce a filter transient as in Remark~\ref{rm:filter-realization}.
If the generator output is identically zero, then $u^\star=0$ and, for example, $\dot u^\star=0$ is already an exact relation.
\end{remark}

Define the forward permanently unobservable subspace
\begin{equation}\label{eq:forward-unobservable-subspace}
\begin{aligned}
\mathcal N_+(t,\mu)=\{x\in\mathbb R^{n_g}:{}&\ell(s,\mu)\Phi_{A_g}(s,t;\mu)x=0\\
&\text{for every }s\ge t\}.
\end{aligned}
\end{equation}
Let $P_o(t,\mu)$ be the orthogonal projection onto $\mathcal N_+(t,\mu)^\perp$.
Its dimension $(n_g-\dim\mathcal N_+(t,\mu))$ counts the output-relevant directions; the generator may have other states.

\begin{proposition}[General LTV systems]\label{prop:ltv-relation}
Suppose $n_g-\dim\mathcal N_+(t,\mu)=r_o\ge1$ for all $t\ge0$ and $\mu\in\mathcal P$.
Suppose there are constants $M_{\rm an},R_{\rm an},\Delta_o,\beta_o>0$, independent of $t,\mu$, such that
\begin{equation}\label{eq:uniform-time-analyticity}
\sup_{t\ge0,\,\mu\in\mathcal P}
\bigl(\|\partial_t^j A_g(t,\mu)\|+\|\partial_t^j\ell(t,\mu)\|\bigr)
\le M_{\rm an}R_{\rm an}^j j!
\end{equation}
for every integer $j\ge0$.
Assume that the forward-window observability Gramian
\begin{equation}\label{eq:observable-window-gramian}
\begin{aligned}
W_o(t,\mu)=\int_0^{\Delta_o}
&\Phi_{A_g}(t+s,t;\mu)^T\ell(t+s,\mu)^T\\
&\cdot\ell(t+s,\mu)\Phi_{A_g}(t+s,t;\mu)\,ds
\end{aligned}
\end{equation}
satisfies, for every $t\ge0$ and $\mu\in\mathcal P$,
\begin{equation}\label{eq:observable-window-bound}
\ker W_o(t,\mu)=\mathcal N_+(t,\mu),\qquad
W_o(t,\mu)\succeq\beta_oP_o(t,\mu).
\end{equation}
Then there exist a common finite $m\ge1$ and $\gamma_J>0$ satisfying~\eqref{eq:finite-derivative-certificate}.
\end{proposition}
\begin{proof}
If no finite derivative prefix satisfied a uniform lower bound on the output-relevant subspace, the uniform analyticity estimates would yield a limiting forward output whose initial derivatives all vanish.
Analyticity forces this output to vanish on the observation window, contradicting the uniform Gramian lower bound.
The finite-order condition then follows from the common output-relevant dimension.
Details are given in Supplementary~S8.1.
\end{proof}

\begin{remark}\label{rm:ltv-relation-conditions}
The factorial bound~\eqref{eq:uniform-time-analyticity} gives the time Taylor expansions a common analytic scale, which is stronger than $C^\infty$ smoothness.
In particular, a generator output whose derivatives all vanish at one instant must vanish thereafter.
Condition~\eqref{eq:observable-window-bound} allows output-irrelevant states and requires uniform observability only on $\mathcal N_+^\perp$.
When $\mathcal N_+=\{0\}$, the generator is uniformly completely observable.
\end{remark}

For periodic systems, compactness gives an alternative sufficient condition without the uniform factorial bound~\eqref{eq:uniform-time-analyticity}.
Periodicity is imposed on the coefficients; individual output trajectories need not be periodic.
\begin{proposition}[Linear periodic systems]\label{prop:periodic-relation}
Suppose $A_g(t,\mu)$ and $\ell(t,\mu)$ have a common period $T_{\mathrm p}>0$, and $\mathcal P$ is compact.
For every fixed $\mu$, suppose these coefficients are real analytic in time, and that each finite-order time derivative is jointly continuous in $(t,\mu)$.
If $n_g-\dim\mathcal N_+(t,\mu)=r_o\ge1$ throughout the family, then there exist a common finite $m\ge1$ and $\gamma_J>0$ satisfying~\eqref{eq:finite-derivative-certificate}.
The coefficients $\delta_J,\beta_j$ in~\eqref{eq:gram-polynomial} have period $T_{\mathrm p}$; the $a_j$ in~\eqref{eq:gram-input-coefficients} also have this period if $b$ does.
\end{proposition}
\begin{proof}
Time analyticity identifies $\mathcal N_+(t,\mu)$ with the common kernel of all output derivative rows.
A finite-cover argument on $[0,T_{\mathrm p}]\times\mathcal P$ gives a common finite prefix of rank $r_o$; continuity and constant rank then give a uniform positive lower bound on its nonzero singular values.
The details are given in Supplementary~S8.2.
\end{proof}

\begin{remark}\label{rm:relation-construction-scope}
Propositions~\ref{prop:ltv-relation} and~\ref{prop:periodic-relation} establish the existence of a finite order $m$ and constants $r_o,\gamma_J$ satisfying~\eqref{eq:finite-derivative-certificate}; once these quantities are verified,~\eqref{eq:gram-polynomial}--\eqref{eq:gram-input-coefficients} give explicit coefficients, although the resulting differential-relation order need not be minimal.
The construction uses polynomial operations on the BI and exosystem coefficients and finitely many of their time derivatives, avoids matrix inversion, and preserves finite separability, when present, after a finite parameter lifting.
A relation can also be obtained directly without the fixed output-relevant dimension assumed in these propositions.
For example, the sum $v$ of the first coordinates of two oscillators with unknown constant frequencies $\mu_1,\mu_2$ satisfies $(\mathscr D^2+\mu_1^2)(\mathscr D^2+\mu_2^2)v=0$.
Here $(n_g-\dim\mathcal N_+)$ equals four for distinct positive frequencies and two when they coincide.
Thus a parameter set containing both cases violates the fixed output-relevant dimension assumed in Propositions~\ref{prop:ltv-relation} and~\ref{prop:periodic-relation}, although the same differential relation remains valid.
\end{remark}

\section{Constructive Designs for Exact and Practical Regulation}
\label{sec:constructive}
This section combines Sections~\ref{sec:robust-regulation} and~\ref{sec:closure} to construct robust regulators from the original system coefficients.
Section~\ref{sec:exact-construction} treats exact regulation under affine parametric uncertainty, while Section~\ref{sec:approximation-construction} develops coefficient approximations for practical regulation.
We consider relative degree one; the higher-relative-degree results and their additional conditions are summarized in Section~\ref{sec:higher-relative-degree}.

\subsection{Exact Regulation Under Affine Parametric Uncertainty}
\label{sec:exact-construction}
We say that the BI and exosystem coefficients $S(t,\mu)$, $Q(t,\mu)$, $p_i(t,\mu)$, $\alpha(t,\mu)$, $\gamma(t,\mu)$, $A_\eta(t,\mu)$, $\beta(t,\mu)$, $P_z(t,\mu)$, and $b(t,\mu)$ have \emph{affine parametric uncertainty} (APU) if they are finitely separable as in Definition~\ref{def:finite-separability}, with continuous parameter functions.
Collecting these functions gives the common representation
\begin{equation}\label{eq:bi-parametric-dictionary}
Y(t,\mu)=Y_0(t)+\sum_{i=1}^{p_a}\vartheta_i(\mu)Y_i(t),
\end{equation}
where $Y$ denotes any of these coefficients and the time functions $Y_i$ are known.
Here ``affine'' refers to the dependence on the finite parameter vector $\vartheta(\mu)$.
This structure is verified in BI coordinates: affine dependence in the original coordinates need not be preserved by a parameter-dependent BI transformation.

Assume that the finite-order condition~\eqref{eq:finite-derivative-certificate} has been verified.
Propositions~\ref{prop:ltv-relation} and~\ref{prop:periodic-relation} provide sufficient conditions for such $m,r_o,\gamma_J$ to exist.
The parameter structure~\eqref{eq:bi-parametric-dictionary} and this finite-order condition are separate requirements.
By Remark~\ref{rm:relation-construction-scope} and~\eqref{eq:rearranged-coefficients}--\eqref{eq:filter-polynomial}, collecting finitely many products of $\vartheta_i(\mu)$ gives a lifting $\theta^\star=T(\mu)$ for which $d$ and every $h_j$ are affine in $\theta^\star$.
The construction retains $v=bu^\star$ and $d=a_m=\delta_Jb^2\ge a_{\min}:=\gamma_J^{r_o}b_{\min}^2$, without parameter-dependent inversion.
The resulting channels define~\eqref{eq:parallel-filter}.

We give a safe-set construction that also applies to the approximations below.
For a fixed $K$, let $T_K:\mathcal P\to\mathbb R^{N_K}$ be a known continuous comparison map with $d_K(t,T_K(\mu))\ge3a_{\min}/4$.
Choose $M_{d\theta,K}\ge\sup_{t\ge0}\|(d_{1,K}(t),\ldots,d_{N_K,K}(t))\|_1$ and finitely many nodes $\mu_l\in\mathcal P$ with a verified covering radius $\rho_{\Theta,K}$ such that
\begin{equation}\label{eq:safe-cover}
\begin{gathered}
\sup_{\mu\in\mathcal P}\min_l
\|T_K(\mu)-T_K(\mu_l)\|_\infty\le\rho_{\Theta,K},\\
M_{d\theta,K}\rho_{\Theta,K}\le a_{\min}/4.
\end{gathered}
\end{equation}
Then use the finite polytope
\begin{equation}\label{eq:certified-safe-set}
\Theta_K=\operatorname{co}\{T_K(\mu_l)\}_l
+\rho_{\Theta,K}[-1,1]^{N_K}.
\end{equation}
The set is nonempty, compact, and convex, contains $T_K(\mathcal P)$, and satisfies $d_K(t,\theta)\ge a_{\min}/2$ throughout $\Theta_K$.
Supplementary~S9 gives its finite constraint description and verifies the covering and denominator bounds using a known continuity modulus or derivative bound for $T_K$.
A directly verified safe polytope may also be used.
If the lifting has no adjustable coordinates, the base filter remains and the parameter update is omitted.

\begin{theorem}[Exact regulation by parameter lifting]\label{thm:exact-construction}
Suppose Assumption~\ref{ass:standing} holds with $r=1$, the BI and exosystem coefficients have APU as defined above, and~\eqref{eq:finite-derivative-certificate} holds for fixed $m,r_o,\gamma_J$.
Use the finite lifting and parallel realization above, with a verified candidate set containing $T(\mathcal P)$ and satisfying $d\ge a_{\min}/2$.
For any $\Gamma>0$, choose $k$ according to~\eqref{eq:r1-gain-conditions} and set $\delta=0$ in~\eqref{eq:r1-controller}.
For every admissible physical parameter and initial state, arbitrary finite actual filter initial states, and $\hat\theta(t_0)\in\Theta$, the closed-loop solution exists uniquely for all $t\ge t_0$ and its states and input are bounded.
Moreover, all three convergence statements in~\eqref{eq:r1-exact-convergence} hold.
\end{theorem}
\begin{proof}
The parallel realization and the verified candidate set give Assumption~\ref{ass:finite-model}(i)--(ii).
The Gram relation has zero persistent residual.
Assigning the compatible ideal initial value from Supplementary~S1.2 to the base block and zero initial values to the other blocks gives the exact ideal filter with zero transient.
Other bounded ideal initial values add only an exponentially decaying, finite-energy homogeneous term, verifying item~(iii).
Theorem~\ref{thm:fixed-model-regulation}, with~\eqref{eq:r1-gain-conditions} and $\delta=0$, proves the stated conclusions for the allowed actual initial conditions.
\end{proof}

\begin{remark}\label{rm:direct-exact-relation}
A directly obtained relation gives the same conclusion when its finite separability and safe candidate set are verified separately; Section~\ref{sec:pendulum} uses such a relation.
Under Assumption~\ref{ass:standing}, the proposed controller also achieves exact regulation when $A,B,C,P,Q$ are uncertain constant matrices and $S(t)$ is known, real analytic, and periodic; see Supplementary~S9.6.
\end{remark}

\subsection{Practical Regulation With a Fixed Stabilization Design}
\label{sec:approximation-construction}
When a nonmonic relation is available but its coefficients are not finitely separable, coefficient approximation provides a route to practical regulation.
Other choices of the approximation target are possible; for example, one may approximate the ideal input $u^\star$ directly, as discussed in Remark~\ref{rm:direct-input-approximation}.

Fix a relation in Definition~\ref{def:nonmonic}, with order $m$ and leading-coefficient bound $a_{\min}$, and fix the Hurwitz template $F_0,C_0,B_j$ in~\eqref{eq:filter-transfer}.
For the reordered scalar coefficients, write
\begin{equation*}
\begin{gathered}
\mathcal C(\mu)=\bigl(d(\cdot,\mu),h_0(\cdot,\mu),\ldots,h_{m-1}(\cdot,\mu)\bigr),\\
\mathcal C(\mu)\in\mathcal X:=(C_b^1)^{m+1},
\end{gathered}
\end{equation*}
where the norm on $\mathcal X$ is the maximum of its component $C_b^1$ norms, and assume $\sup_{\mu\in\mathcal P}\|\mathcal C(\mu)\|_{\mathcal X}<\infty$.
Use affine approximations $\mathcal C_K(\theta)=(d_K(\cdot,\theta),h^{\rm rel}_{0,K}(\cdot,\theta),\ldots,h^{\rm rel}_{m-1,K}(\cdot,\theta))$ with constant comparison parameters $\theta_K^\star=T_K(\mu)$ satisfying
\begin{equation}\label{eq:coefficient-c1-approx}
\sup_{\mu\in\mathcal P}
\|\mathcal C_K(T_K(\mu))-\mathcal C(\mu)\|_{\mathcal X}
\le\eta_K,\; \eta_K\longrightarrow0.
\end{equation}
Each fixed $K$ has known coefficient channels and an implementable compact convex candidate set containing $T_K(\mathcal P)$, with $d_K\ge d_{\min}>0$ throughout that set and $d_{\min}$ independent of $K$.
The scalar coefficients $h^{\rm rel}_{j,K}$ are those of Remark~\ref{rm:coefficient-residual}, distinct from the stacked readout row $h_K$.
The $C_b^1$ estimate also controls the derivative term in the effective cascade below.

Write $h^{\rm rel}_{j,K}(t,\theta)=h^{\rm rel}_{j0,K}(t)+\sum_{i=1}^{N_K}\theta_i h^{\rm rel}_{ji,K}(t)$ and use the parallel blocks with $G_{i,K}=\sum_{j=0}^{m-1}B_jh^{\rm rel}_{ji,K}$.
Let $c_{0,K}^\star=1$, $c_{i,K}^\star=\theta_{i,K}^\star$ for $i\ge1$, and define the analysis combinations
\begin{equation}\label{eq:effective-combinations}
\begin{aligned}
\bar\zeta_K&=\sum_{i=0}^{N_K}c_{i,K}^\star\zeta_{i,K},&
\bar G_K&=\sum_{i=0}^{N_K}c_{i,K}^\star G_{i,K},\\
\bar R_K&=\bar G_K/b,&
\bar q_K&=\bar\zeta_K-\bar\zeta_K^\star-\bar R_Ke.
\end{aligned}
\end{equation}
Here $\bar\zeta_K^\star$ is the same combination of the ideal filter states.
The weights are constant comparison parameters and may have either sign.
For a representation without a base block, the sums start at $i=1$.
The identity
\begin{equation}\label{eq:effective-forcing}
\bar G_K(t,\mu)=\sum_{j=0}^{m-1}B_j
h^{\rm rel}_{j,K}(t,T_K(\mu))
\end{equation}
gives common bounds on $\bar G_K$ and $\dot{\bar G}_K$ directly from~\eqref{eq:coefficient-c1-approx}, with no factor depending on $N_K$.

\begin{lemma}[Common realization bounds]\label{lem:effective-realization-bounds}
Under Assumption~\ref{ass:standing} with $r=1$ and the coefficient and candidate-set conditions above, the parallel realizations satisfy Assumption~\ref{ass:finite-model} for each fixed $K$, with
\begin{equation}\label{eq:constructive-residual}
\varepsilon_K=M_{u^\star}M_Q\eta_K.
\end{equation}
They also satisfy the common comparison estimates~\eqref{eq:family-comparison-bounds}.
In particular, the effective state $x_{\mathrm{eff},K}=\operatorname{col}(z,\bar q_K)$ has dimension $(n-1+m)$ and satisfies
\begin{equation}\label{eq:effective-cascade}
\begin{aligned}
\dot z&=A_\eta z+\beta e,\\
\dot{\bar q}_K&=F_0\bar q_K-\bar R_K\gamma z\\
&\quad+(F_0\bar R_K-\dot{\bar R}_K-\bar R_K\alpha)e.
\end{aligned}
\end{equation}
The homogeneous cascade has exponential stability constants independent of $K$.
Its Lyapunov norm $W_K$ bounds the actual error coupling, since
\begin{equation}\label{eq:effective-matching}
\begin{aligned}
p_K&=d_K^\star/(\sigma_b b),\qquad d_K^\star=d_K(t,T_K(\mu)),\\
A_{e,K}&=p_K\alpha+C_0\bar G_K/(\sigma_b b),\\
B_{e,K}\chi_K&=p_K\gamma z+\sigma_b C_0\bar q_K.
\end{aligned}
\end{equation}
The common constants depend only on the plant bounds, the fixed filter template, the comparison coefficient bounds, and $d_{\min}$.
\end{lemma}
\begin{proof}
The input difference cancels because $\bar R_Kb=\bar G_K$.
Appendix~\ref{app:effective-realization-bounds} proves the effective cascade and common comparison bounds; explicit constants and residual and initialization details are given in Supplementary~S9.3.
\end{proof}

\begin{remark}\label{rm:effective-state-bounds}
Once the regulation estimate bounds $e_K$,~\eqref{eq:family-full-filter-recovery} recovers every filter state for that fixed $K$, followed by the readout, input, regressor, and update rate.
Thus the lemma does not require common Euclidean bounds on the full stacked matrices or actual state peaks.
Candidate-set diameters, comparison initial values, parameter storage, and transient energies may also depend on $K$, subject to the fixed-realization bounds in Assumption~\ref{ass:finite-model}.
\end{remark}

\begin{corollary}[Practical regulation by approximation]\label{cor:constructive-practical}
Suppose Assumption~\ref{ass:standing} holds with $r=1$, and the affine coefficient approximations and candidate sets satisfy the conditions of Lemma~\ref{lem:effective-realization-bounds}.
Index the resulting family by $K\ge K_0$ and choose $\Gamma_K=\gamma_{\rm ad}I_{N_K}$ with a fixed $\gamma_{\rm ad}>0$.
Choose one $k$ satisfying~\eqref{eq:r1-gain-conditions} with the common constants of Lemma~\ref{lem:effective-realization-bounds}, and let $\kappa>0$ be the resulting margin.
With $\varepsilon_K$ from~\eqref{eq:constructive-residual} and $\delta_K=2\varepsilon_K/\kappa$, controller~\eqref{eq:r1-controller} has bounded states and input and a unique global solution for each fixed $K$, every admissible plant and exosystem initial state, finite actual filter initial state, and estimate initialized in $\Theta_K$.
Moreover,
\begin{equation}\label{eq:constructive-practical}
\begin{gathered}
\limsup_{t\to\infty}|e_K(t)|\le\frac{2M_{u^\star}M_Q}{\kappa}\eta_K,\\
\lim_{K\to\infty}\limsup_{t\to\infty}|e_K(t)|=0.
\end{gathered}
\end{equation}
The stabilization gain, relation order, and filter template are fixed before the finite realization is selected offline.
\end{corollary}
\begin{proof}
Lemma~\ref{lem:effective-realization-bounds} verifies Assumption~\ref{ass:model-family} with $\varepsilon_K=M_{u^\star}M_Q\eta_K\to0$ and finite transient energy for each fixed $K$.
Theorem~\ref{thm:model-family-regulation} applies with the same $k$ and $\delta_K=2\varepsilon_K/\kappa$.
Equation~\eqref{eq:family-full-filter-recovery} bounds every actual filter state, including those absent from the effective state.
Taking the time limit for each fixed finite $K$ first and then the limit in $K$ proves~\eqref{eq:constructive-practical}.
\end{proof}
In contrast to the high-gain route in Remark~\ref{rm:high-gain-comparison}, this construction improves the error bound by refining the coefficient approximation with a fixed stabilization gain.
The realization dimension remains fixed during each run; neither parameter convergence nor a common learning time is required.
\begin{remark}[Weighted Taylor approximation]\label{rm:taylor-approximation}
Different coefficient approximations satisfying~\eqref{eq:coefficient-c1-approx} can be used in Corollary~\ref{cor:constructive-practical}, provided that their candidate sets satisfy the conditions of Lemma~\ref{lem:effective-realization-bounds}.
A weighted Taylor expansion gives one such construction under the following analyticity conditions.
Suppose $\mathcal P\subset\{\mu:|\mu_i-\mu_{0,i}|\le r_i\}$ and $\mathcal C$ extends holomorphically to a neighborhood of the closed complex polydisc $|\mu_i-\mu_{0,i}|\le R_i$, where $0<r_i<R_i$.
Assume its $\mathcal X$ norm is bounded there by $M_{\mathcal C}$; this includes a uniform bound for the first time derivatives.
For a multi-index $\alpha\in\mathbb N_0^{n_\mu}$, write $\mathcal C_\alpha=(\partial_\mu^\alpha\mathcal C)(\mu_0)/\alpha!$.
Choose $r_i<\rho_i<R_i$, and set
\begin{equation}\label{eq:taylor-coefficients}
\begin{aligned}
\theta_\alpha^\star&=(\mu-\mu_0)^\alpha/\rho^\alpha,
&\bar{\mathcal C}_\alpha&=\rho^\alpha\mathcal C_\alpha,\\
\mathcal C_K(\theta)&=\mathcal C_0+
\sum_{1\le|\alpha|\le K}\theta_\alpha\bar{\mathcal C}_\alpha,
\end{aligned}
\end{equation}
The remainder estimate in Supplementary~S9 gives
\begin{equation}\label{eq:taylor-c1-bound}
\eta_K\le M_{\mathcal C}\sum_{n>K}
\binom{n+n_\mu-1}{n_\mu-1}q^n,
\end{equation}
where $q=\max_i(r_i/R_i)<1$.
The series tail tends to zero as $K\to\infty$.
The scaling $\rho$ adjusts the numerical sizes of the coefficients and parameters; Section~\ref{sec:motor} uses its scalar version.

For all sufficiently large $K$, choose the bound $\eta_K\le a_{\min}/4$.
Then $d_K(t,T_K(\mu))\ge3a_{\min}/4$, and~\eqref{eq:safe-cover}--\eqref{eq:certified-safe-set} give $d_{\min}=a_{\min}/2$.
A larger independently varying parameter box requires separate verification of its denominator bound.

For the Gram construction in Section~\ref{sec:closure}, a sufficient condition for the stated parameter analyticity is that the original BI coefficients have a common bounded holomorphic extension in the finite time-derivative norms used in the construction.
Finite differentiation and polynomial operations then transfer the property to $\mathcal C$.
Stability is required only for real physical parameters; the polynomial Gram formulas extend using the algebraic transpose.
\end{remark}

\begin{remark}[Approximation by convex interpolation]\label{rm:convex-interpolation}
Convex interpolation provides an alternative for coefficient approximation; details are given in Supplementary~S9.5.
Suppose $\mathcal C:\mathcal P\to\mathcal X$ is continuous on the compact parameter set.
For nodes $\mu_i$ forming an $\ell^\infty$-cover of radius $h_N\to0$, use $\mathcal C_N(\theta)=\sum_{i=1}^N\theta_i\mathcal C(\mu_i)$ with $\theta_i\ge0$ and $\sum_i\theta_i=1$.
Constant comparison weights give an approximation error at most $\omega_{\mathcal C}(h_N)\to0$, where $\omega_{\mathcal C}$ is the modulus of continuity in $\mathcal X$.
The candidate simplex preserves the common node bound $d(t,\mu_i)\ge a_{\min}$.
Thus Corollary~\ref{cor:constructive-practical} applies without parameter analyticity.
The node coefficients are known, and the controller estimates their weights directly.
\end{remark}

\begin{remark}\label{rm:direct-input-approximation}
As an alternative to using a differential relation, adaptive regulation can be based directly on finite separation or approximation of the ideal input.
For $r=1$, a representation $u^\star=\psi^T\alpha^\star+\tau$ with $\tau\in L^2\cap L^\infty$ gives a direct adaptive feedforward design achieving exact regulation.
The same design treats uniform finite approximations over the uncertainty set and the entire forward interval, with a feedback gain independent of the number of terms; see Supplementary~S10.
The allowance of general vanishing remainders in Supplementary~S3.3 concerns the earlier linear design in~\cite[Corollary~4.1]{zha2026robust}, not this adaptive feedforward law.
However, such uniform signal approximations may fail even when the relation coefficients are finitely separable: for $\mu\in[1,2]$, $u_\mu^\star(t)=\sin(\mu t)$ admits no arbitrarily accurate uniform finite-dimensional approximation on $[0,\infty)$ by known bounded time functions, even with the stated transients, whereas $\ddot u_\mu^\star+\mu^2u_\mu^\star=0$ has finitely separable coefficients.
See Supplementary~S10 for the proof.
\end{remark}

\section{Illustrative Examples}\label{sec:examples}
All simulations are performed in MATLAB/Simulink.
\subsection{Exact Regulation of a Pendulum System}\label{sec:pendulum}
Consider a pendulum model adapted from~\cite{zhang2009tac}, with the composite output $y=\xi=\dot z_p+\lambda z_p$ and reference $(w_1+w_2)$:
\begin{equation}\label{eq:pendulum-model}
\begin{aligned}
\dot\xi&=\lambda\xi-(a_p+\lambda^2)z_p+u,\;\dot z_p=\xi-\lambda z_p,\\
\dot w_1&=w_2,\quad\dot w_2=-(a_p+2d_p\cos2t)w_1,\\
e&=\xi-w_1-w_2,
\end{aligned}
\end{equation}
where $a_p=1.6$, $d_p=0.1$, and $\lambda\in[0.75,1.25]$ is unknown.
The plant has relative degree one and the stable zero dynamics.
The ideal-input family, as $\lambda$ and the admissible exosystem initial state vary, has infinite-dimensional span modulo terms tending to zero, by an argument similar to that in~\cite[Example~3]{zha2026robust}.
Thus this family cannot be generated asymptotically by a fixed finite-dimensional linear internal model in zero-input operation~\cite[Theorem~4.1 and Corollary~4.1]{zha2026robust}.
However, the adaptive design developed here overcomes this limitation.

\begin{figure}[!t]
\centering
\includegraphics[width=3.15in]{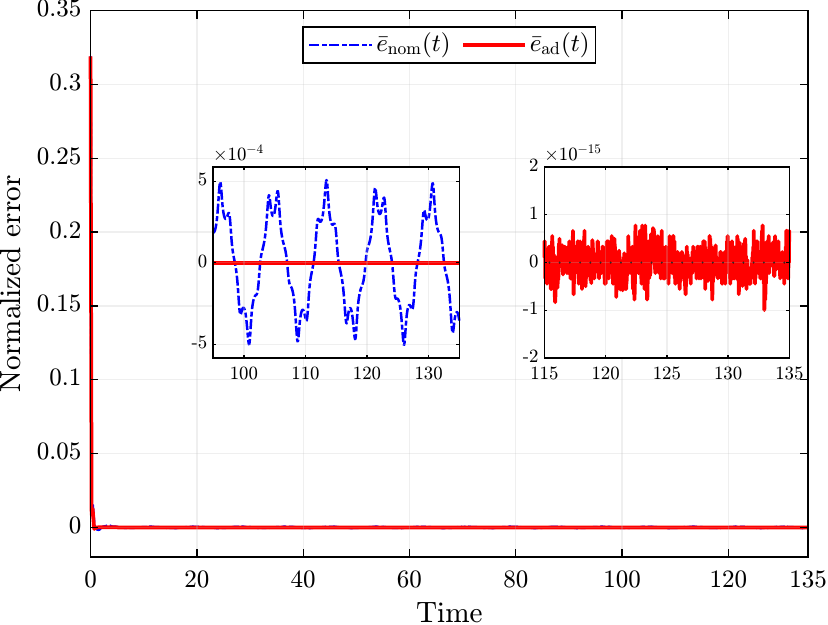}
\caption{Normalized regulation errors of the nominal and adaptive controllers for the pendulum system.}
\label{fig:pendulum-error}
\end{figure}

Set $f(t)=2d_p\cos2t$ and let $\Delta_p,\beta_0,\beta_1$ be the coefficients constructed in Supplementary~S11.1.
Since $\Delta_p=2(\dot f+f/2)^2+(a_p+9/2+f)f^2>0$, periodicity and the absence of simultaneous zeros of $f$ and $\dot f$ give a uniform positive lower bound.
The resulting exact third-order relation is
\begin{equation}\label{eq:pendulum-relation}
\begin{aligned}
&\Delta_p u^{\star(3)}+(\lambda\Delta_p-\beta_1)\ddot u^\star\\
&\qquad-(\lambda\beta_1+\beta_0)\dot u^\star-\lambda\beta_0u^\star=0.
\end{aligned}
\end{equation}
The coefficients of~\eqref{eq:pendulum-relation} are affine in $\lambda$.
Applying~\eqref{eq:rearranged-coefficients}--\eqref{eq:filter-polynomial} and~\eqref{eq:parallel-filter} with $\mathsf P(s)=(s+3)^3$ gives two three-dimensional stable filter blocks.
With $C_0=[0,0,1]$, their readout and feedback are $u_{\rm im}=(C_0\zeta_0+\hat\lambda C_0\zeta_1)/\Delta_p$ and $u=u_{\rm im}-20e$.
We use~\eqref{eq:r1-controller} with $\delta=0$, adaptation gain $\gamma_\lambda=10^7$, and projection onto $[0.75,1.25]$.

Set $\lambda=0.75$, $\hat\lambda(0)=1$, $\xi(0)=z_p(0)=0$, and $w(0)=[0.02,-0.04]^T$, with all six filter states initially zero.
The adaptive controller updates $\hat\lambda$ throughout the simulation.
For comparison, the nominal controller uses the same settings, except that $\hat\lambda$ is frozen at $1$.
The simulation horizon is $[0,400]$.

To evaluate the results, we use the normalized error $\bar e=e/\phi_r$ hereafter, where $\phi_r$ is the numerical estimate of the peak absolute reference over the corresponding simulation horizon.
Fig.~\ref{fig:pendulum-error} compares the normalized errors of the nominal and adaptive controllers.
Over $[300,400]$, their normalized peak absolute errors are $5.08\times10^{-4}$ and $1.22\times10^{-15}$, respectively; the latter is at the numerical resolution level of the simulation.
The parameter estimate approaches the true value in this run (see Fig.~S1), although parameter convergence is not asserted by Theorem~\ref{thm:fixed-model-regulation}.

\subsection{Practical Regulation of a DC Motor}\label{sec:motor}
Consider current regulation of a DC motor subject to an unknown load torque $w_1$:
\begin{equation*}
L_a\dot i=-R_ai-K_e\omega+V,\qquad
J_m\dot\omega=K_ti-B_m\omega-w_1.
\end{equation*}
We set the constant load bias $\tau_b=0$ and choose the simulation parameters $R_a=0.30\,\Omega$, $L_a=10^{-4}\,\mathrm H$, $K_t=0.030\,\mathrm{N\,m/A}$, $K_e=0.030\,\mathrm{V\,s/rad}$, $J_m=10^{-4}\,\mathrm{kg\,m^2}$, and $B_m=10^{-4}\,\mathrm{N\,m\,s/rad}$.
The target is $i_{\rm ref}=1\,\mathrm A$, equivalently a constant electromagnetic torque of $0.030\,\mathrm{N\,m}$; the speed is allowed to vary with the load.
With $\xi=i-1$, $z_p=\omega-300$, and $u=V-9.3$, the BI form is
\begin{equation}\label{eq:motor-bi}
\begin{aligned}
\dot\xi&=-3000\xi-300z_p+10000u,\\
\dot z_p&=300\xi-z_p-10000w_1,\qquad e=\xi.
\end{aligned}
\end{equation}
It has relative degree one and stable zero dynamics.
\begin{figure}[!t]
\centering
\includegraphics[width=3.15in]{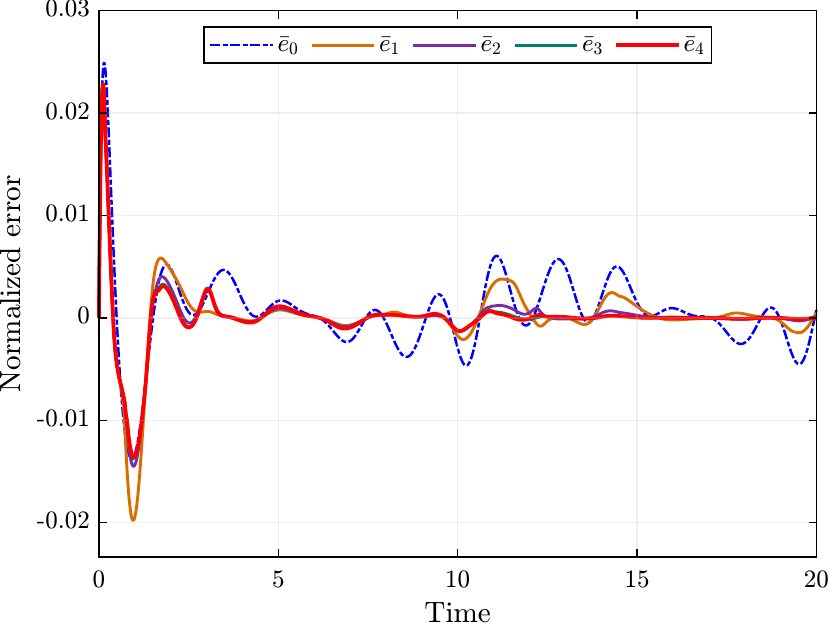}
\caption{Transient normalized errors $\bar e_K$ for Taylor truncation degrees $K=0,1,2,3,4$.}
\label{fig:motor-transient}
\end{figure}

\begin{figure}[!t]
\centering
\includegraphics[width=3.5in]{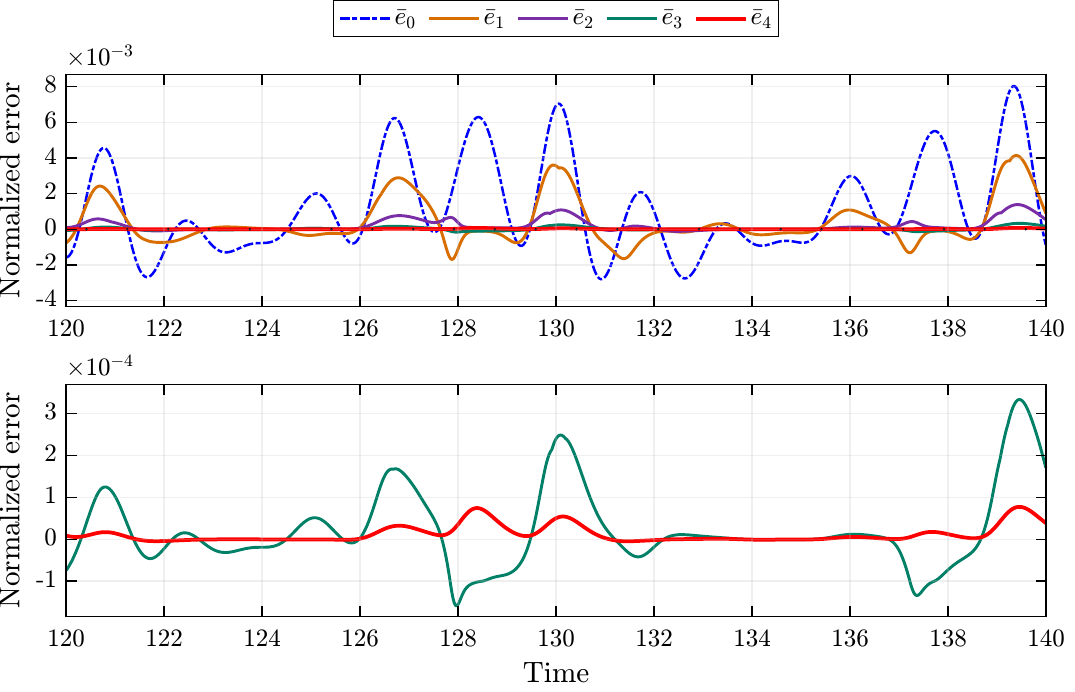}
\caption{Late-time normalized errors for the DC motor.
The lower panel shows a zoomed-in view of the errors for $K=3,4$ on a finer vertical scale.}
\label{fig:motor-late}
\end{figure}

The load is generated by the prescribed exosystem family
\begin{equation*}
\begin{gathered}
\dot w=\nu(t,\mu)\begin{bmatrix}0&1\\-1&0\end{bmatrix}w,
\qquad \nu(t,\mu)=2e^{\mu h(t)},\\
h(t)=\tfrac12\bigl(\sin2t+\sin(\sqrt2t)\bigr),\qquad
\mu\in[-0.25,0.25].
\end{gathered}
\end{equation*}
Its skew-symmetric matrix preserves $\|w\|$, while the incommensurate frequencies give nonperiodic coefficients for $\mu\ne0$.

To construct the input relation directly from~\eqref{eq:motor-bi}, put $a_{\omega}=B_m/J_m=1$ and use the generator and output derivative matrix $J_3$ in Supplementary~S11.2.
Since $\det J_3=\nu$ and $2e^{-1/4}\le\nu\le2e^{1/4}$, both $J_3$ and $J_3^{-1}$ are uniformly bounded, which verifies the finite-order rank and uniform nondegeneracy conditions in~\eqref{eq:finite-derivative-certificate}.
The Gram construction gives
\begin{equation}\label{eq:motor-relation}
\begin{aligned}
&\nu^2u^{\star(3)}+(a_{\omega}\nu^2-\nu\dot\nu)\ddot u^\star\\
&\qquad+(\nu^4-a_{\omega}\nu\dot\nu)\dot u^\star+a_{\omega}\nu^4u^\star=0.
\end{aligned}
\end{equation}
Thus the construction uses the original model coefficients without first explicitly solving the regulator equations.

For $0\le i\le3$, let $a_i(t,\mu)$ denote the coefficient of $u^{\star(i)}$ in~\eqref{eq:motor-relation}, and set $a_{ij}(t)=(1/j!)\partial_\mu^j a_i(t,0)$ for $j\ge0$.
Their explicit expressions are given in Supplementary~S11.2.
For truncation degree $K$, use $\rho=0.5$, $\bar a_{ij}=\rho^j a_{ij}$, and $\theta_j^\star=(\mu/\rho)^j$ to define
\begin{equation}\label{eq:motor-taylor}
\begin{gathered}
a_{i,K}(t,\theta)=a_{i0}(t)+\sum_{j=1}^K\theta_j\bar a_{ij}(t),\\
\Theta_K=\prod_{j=1}^K[-2^{-j},2^{-j}].
\end{gathered}
\end{equation}
The leading coefficient $d_K(t,\theta)=4+\sum_{j=1}^K\theta_jd_j(t)$, with $d_j=4\rho^j(2h)^j/j!$, satisfies $d_K(t,\theta)\ge4(2-e^{1/2})>0$ for every $\theta\in\Theta_K$.
Apply~\eqref{eq:rearranged-coefficients}--\eqref{eq:filter-polynomial} to the channels in~\eqref{eq:motor-taylor}, with $\mathsf P(s)=(s+4)^3$, and use~\eqref{eq:parallel-filter}.
The controller~\eqref{eq:r1-controller} has
\begin{equation*}
\begin{aligned}
u_{\rm im}&=\frac{C_0\zeta_0+\sum_{j=1}^K\hat\theta_jC_0\zeta_j}
{d_K(t,\hat\theta)},\qquad u=u_{\rm im}-5e,\\
\phi_j&=d_ju-C_0\zeta_j,\qquad \Gamma_K=2000I_K,
\end{aligned}
\end{equation*}
with the positive-sign projected update and $\delta_K=10^{-4}\,\mathrm A$.
The feedback gain $k=5\,\mathrm{V/A}$ and the dead-zone width are chosen numerically.
For $K=0$, the controller still contains a three-state nominal internal model, with $d_0=4$, but no adaptive coordinates.
For $K\ge1$, it has $3(K+1)$ filter states and $K$ parameter estimates.
Here $K$ is the Taylor truncation degree; the plant relative degree and the differential relation order remain one and three, respectively.

We compare $K=0,\ldots,4$ at $\mu=0.25$.
All cases use the same settings, except for the additional filter channels and parameter coordinates prescribed by~\eqref{eq:motor-taylor}.
Set $i(0)=1\,\mathrm A$, $\omega(0)=300\,\mathrm{rad/s}$, and $w(0)=[0.008,0]^T\,\mathrm{N\,m}$, with load bound $\|w\|\le0.010\,\mathrm{N\,m}$ and zero initial filter states and parameter estimates.
The horizon is $[0,180]\,\mathrm s$, with $\phi_r=1\,\mathrm A$.

Figs.~\ref{fig:motor-transient} and~\ref{fig:motor-late} show the initial transient and late-time normalized errors, respectively.
Over $[120,180]\,\mathrm{s}$, the normalized peak absolute error is reduced by a factor of about $104$ from $K=0$ to $K=4$ under the fixed design settings.

\section{Conclusion}
\label{sec:conclusion}
This paper has developed an adaptive output regulation approach for linear time-varying systems with unknown constant parameters in both the plant and exosystem, based on nonmonic differential relations for the ideal input.
Sufficient conditions for the existence of these relations and finite-order constructions have been obtained from the original system coefficients.
The relations lead to a finite-dimensional adaptive internal model consisting of stable filters and a parameterized readout.
Under the stated realization conditions, the resulting controller achieves exact regulation for finitely separable relation coefficients, including the affine parametric uncertainty treated in Byrnes--Isidori form.
For coefficient families satisfying the approximation conditions, improving a finite approximation yields practical regulation while keeping the stabilization design fixed.
Neither persistent excitation nor parameter convergence is required.
Future work will further extend the class of systems for which robust regulation can be achieved.

\appendix
\section{Proofs of the Main Results}\label{app:proofs}
\subsection{Proof of Theorem~\ref{thm:fixed-model-regulation}}
\label{app:fixed-model-regulation}
\begin{proof}
Fix $\mu\in\mathcal P$ and a comparison parameter and trajectory from Assumption~\ref{ass:finite-model}.
The positive denominator and first-derivative bounds make the unprojected closed-loop field locally Lipschitz on a neighborhood of the safe set.
Applying Supplementary~S1.3 to the free states $(x,w,\zeta)$ and the constrained parameter gives a unique local Carath\'eodory solution with $\hat\theta(t)\in\Theta$.
We derive bounds on this solution before proving global continuation.

To separate the stable dynamics from the input mismatch, subtract the filter equations to obtain $\dot{\tilde\zeta}=F\tilde\zeta+G(u-u^\star)$.
The transformation~\eqref{eq:r1-transformation} then gives
\begin{equation*}
\begin{aligned}
\dot\chi&=A_\chi(t,\mu)\chi+B_\chi(t,\mu)e,\\
A_\chi&=\begin{bmatrix}
A_\eta&0\\-R_f\gamma&F
\end{bmatrix},\\
B_\chi&=\operatorname{col}\bigl(\beta,FR_f-\dot R_f-R_f\alpha\bigr).
\end{aligned}
\end{equation*}
Here and below, suppressed arguments in coefficient formulas are $(t,\mu)$ for plant coefficients and $t$ for filter coefficients.
The diagonal systems generated by $A_\eta$ and $F$ are uniformly exponentially stable, and $R_f\gamma$ is uniformly bounded.
The off-diagonal transition block is a convolution bounded by $M_{\mathrm{cas}}(1+t-s)e^{-\nu(t-s)}$, where $\nu>0$ is the smaller diagonal decay rate.
Thus $A_\chi$ is uniformly exponentially stable at any smaller positive rate, with bounds independent of $k$, $\delta$, and $\Gamma$.
The matrix $B_\chi$ is bounded because $b$ is bounded away from zero and $\dot b$ and $\dot G$ are bounded.

Let $\Phi_\chi(s,t;\mu)$ denote the transition matrix of $A_\chi$, and define $P_\chi(t,\mu)=\int_t^\infty\Phi_\chi(s,t;\mu)^T\Phi_\chi(s,t;\mu)\,ds$.
It satisfies $\dot P_\chi+A_\chi^TP_\chi+P_\chi A_\chi=-I$.
Exponential stability bounds this integral from above, while boundedness of $A_\chi$ gives a positive lower bound by integration over a fixed short interval.
Hence, for some $\underline P_\chi,\overline P_\chi>0$,
\begin{equation}\label{eq:r1-matrix-bounds}
\underline P_\chi I\le P_\chi(t,\mu)\le\overline P_\chi I.
\end{equation}
For $W=(\chi^TP_\chi\chi)^{1/2}$, choose $\lambda_\chi=1/(2\overline P_\chi)$ and $M_{\chi e}=\overline P_\chi\|B_\chi\|_\infty/\sqrt{\underline P_\chi}$.
The Lyapunov identity gives~\eqref{eq:r1-comparison-norm} when $W>0$.
At $\chi=0$, the upper right Dini derivative is bounded by $\sqrt{\overline P_\chi}\|B_\chi\|_\infty|e|\le M_{\chi e}|e|$, so the same inequality holds.
Only the resulting uniform scalar bounds are needed to choose the controller gain.

For explicit choices in~\eqref{eq:r1-coefficient-bounds}, let $M_d$, $M_{\partial_t d}$, and $M_h$ bound $d(t,\theta)$, $|\partial_t d(t,\theta)|$, and $\|h(t,\theta)\|$, respectively, for all $t\ge0$ and $\theta\in\Theta$.
Let $M_{\dot b}$ bound $|\dot b(t,\mu)|$ uniformly.
The assumptions make these bounds finite, and one may take
\begin{equation*}
\begin{aligned}
p_{\min}&=\frac{d_{\min}}{M_b},&
M_p&=\frac{M_d}{b_{\min}},\\
M_{\dot p}&=\frac{M_{\partial_t d}}{b_{\min}}
       +\frac{M_dM_{\dot b}}{b_{\min}^2},\\
M_{A_e}&=M_p\|\alpha\|_\infty
       +\frac{M_h\|G\|_\infty}{b_{\min}},\\
M_{B_e}&=\frac{\sqrt{M_p^2\|\gamma\|_\infty^2+M_h^2}}
              {\sqrt{\underline P_\chi}}.
\end{aligned}
\end{equation*}
The coefficient norms include the admissible parameter set, and $M_{\partial_t d}$ bounds $\dot d^\star$ because $\theta^\star$ is constant.
These bounds are independent of $k$, so both inequalities in~\eqref{eq:r1-gain-conditions} hold for a finite choice of $k$.

We next use the affine structure to match the parameter update with the error dynamics.
Affineness and the control law give $d^\star u-h^\star\zeta+\phi^T\tilde\theta=-k\sigma_b\widehat d\,e$.
Subtracting $d^\star u^\star-h^\star\zeta^\star=\Delta_{\mathrm p,K}+\tau_K$ gives~\eqref{eq:r1-matching}.
Multiplying the error equation~\eqref{eq:error-degree-one} by $p=d^\star/(\sigma_b b)$ and substituting $\tilde\zeta=q_f+R_fe$ then gives
\begin{equation}\label{eq:r1-weighted-error}
\begin{aligned}
p\dot e={}&-k\widehat d\,e+A_e e+B_e\chi\\
&-\sigma_b\phi^T\tilde\theta-\sigma_b(\Delta_{\mathrm p,K}+\tau_K).
\end{aligned}
\end{equation}
For the truncated variables in~\eqref{eq:r1-storage}, local absolute continuity gives $E_\delta\dot E_\delta=E_\delta\dot e$ almost everywhere.
The squared truncations are continuously differentiable in their scalar arguments, so the following storage estimates remain valid across dead-zone crossings in the integrated sense.
Since $|e|\le\delta+|E_\delta|$, the comparison estimate for $W$ yields
\begin{equation*}
X_\delta\dot X_\delta
\le-\lambda_\chi X_\delta^2+M_{\chi e}X_\delta|E_\delta|
\end{equation*}
almost everywhere, including the region where $X_\delta=0$.
For $V_\theta=\tfrac12\tilde\theta^T\Gamma^{-1}\tilde\theta$, the update law and~\textup{(S4)} give $\dot V_\theta\le\sigma_b E_\delta\phi^T\tilde\theta$.
Adding this estimate to~\eqref{eq:r1-weighted-error} multiplied by $E_\delta$ cancels the parameter cross term.
Using $eE_\delta=E_\delta^2+\delta|E_\delta|$, $\widehat d\ge d_{\min}$, and $W\le X_\delta+M_{\chi e}\delta/\lambda_\chi$, we obtain
\begin{equation*}
\begin{aligned}
\dot V_\delta\le{}&
-\lambda_\chi X_\delta^2
-\left(kd_{\min}-M_{A_e}-\frac{M_{\dot p}}{2}\right)E_\delta^2\\
&+(M_{\chi e}+M_{B_e})X_\delta|E_\delta|
-(\kappa\delta-\varepsilon_K)|E_\delta|\\
&+|\tau_K||E_\delta|.
\end{aligned}
\end{equation*}
Young's inequality bounds the cross term by
$(\lambda_\chi X_\delta^2/2+(M_{\chi e}+M_{B_e})^2E_\delta^2/(2\lambda_\chi))$ and proves~\eqref{eq:r1-dissipation}.
To prove boundedness for every fixed finite $\delta$, define
\begin{equation*}
M_{V_\theta}=\frac12\sup_{\theta_1,\theta_2\in\Theta}
(\theta_1-\theta_2)^T\Gamma^{-1}(\theta_1-\theta_2).
\end{equation*}
Compactness gives $V_\theta\le M_{V_\theta}<\infty$.
Dropping the nonpositive term $-\kappa\delta|E_\delta|$ in~\eqref{eq:r1-dissipation} and using
$(\varepsilon_K+|\tau_K|)|E_\delta|\le c_eE_\delta^2/2+(\varepsilon_K^2+|\tau_K|^2)/c_e$ gives
\begin{equation*}
\begin{aligned}
\dot V_\delta
&\le-\frac{\lambda_\chi}{2}X_\delta^2
     -\frac{c_e}{2}E_\delta^2+\frac{\varepsilon_K^2+|\tau_K|^2}{c_e}\\
&\le-\lambda_V V_\delta+\lambda_V M_{V_\theta}
       +\frac{\varepsilon_K^2+|\tau_K|^2}{c_e},\\
\lambda_V&=\min\{\lambda_\chi,c_e/M_p\}>0.
\end{aligned}
\end{equation*}
Thus, on the local solution,
\begin{equation*}
\begin{aligned}
V_\delta(t)\le{}&
e^{-\lambda_V(t-t_0)}V_\delta(t_0)\\
&+\left(M_{V_\theta}+\frac{\varepsilon_K^2}{c_e\lambda_V}\right)
\bigl(1-e^{-\lambda_V(t-t_0)}\bigr)\\
&+\frac1{c_e}\int_{t_0}^t e^{-\lambda_V(t-s)}|\tau_K(s)|^2\,ds.
\end{aligned}
\end{equation*}
Since $\tau_K\in L^2$, the last integral is bounded by $\|\tau_K\|_2^2$, so $V_\delta$ is bounded for every fixed finite $\delta$.
Since $p\ge p_{\min}>0$, $|e|\le\delta+|E_\delta|$, and $W\le M_{\chi e}\delta/\lambda_\chi+X_\delta$, both $e$ and $W$ are bounded.
The matrix bound~\eqref{eq:r1-matrix-bounds} then bounds $z$ and $q_f$.
The identity $\zeta=\zeta^\star+q_f+R_fe$ bounds the actual filter state.
The safe denominator and coefficient bounds then bound $\widehat H$ and $u$.
Proposition~\ref{prop:ideal-generator} bounds $w$ and $z_g$, while $\xi_1=e-Qw$, $z_p=z+z_g$, and the bounded inverse BI transformation bound $x$.
Together with $\hat\theta\in\Theta$, these bounds exclude finite-time escape and extend the unique solution to $[t_0,\infty)$.
All these bounds are independent of the maximal existence time.

Now impose~\eqref{eq:r1-dead-zone-condition}.
Using $|\tau_K||E_\delta|\le c_eE_\delta^2/2+|\tau_K|^2/(2c_e)$ in~\eqref{eq:r1-dissipation} gives~\eqref{eq:r1-transient-dissipation}.
Integration yields
\begin{equation*}
\begin{aligned}
V_\delta(t)&+\int_{t_0}^t
\left(\frac{\lambda_\chi}{2}X_\delta(s)^2+\frac{c_e}{2}E_\delta(s)^2\right)\,ds\\
&\le V_\delta(t_0)+\frac1{2c_e}\|\tau_K\|_2^2.
\end{aligned}
\end{equation*}
Thus $X_\delta,E_\delta\in L^2$; the finite-energy supply bounds the storage without requiring it to be monotone.
Boundedness already established above makes $\dot e$ bounded in~\eqref{eq:error-degree-one}.
The Lyapunov identity also gives
$|\dot W|\le W/(2\underline P_\chi)+M_{\chi e}|e|$ almost everywhere.
Consequently, $E_\delta$ and $X_\delta$ are uniformly continuous.
A uniformly continuous square-integrable function must tend to zero: otherwise uniform continuity gives infinitely many disjoint intervals of a common positive length on which its magnitude is bounded away from zero, contradicting square integrability.
Applying this fact to $E_\delta$ and $X_\delta$ gives their convergence to zero.
This proves~\eqref{eq:r1-regulation-bound}.

For zero persistent residual and $\delta=0$, take $\varepsilon_K=0$ in the same estimate, allowing $\tau_K$ to be nonzero.
Then $e,W\to0$, and hence $z,q_f\to0$.
It remains to prove input convergence.
The bounded regressor and~\textup{(S5)} bound $\dot{\hat\theta}$ almost everywhere.
Write $H(t,\theta)=h(t,\theta)/d(t,\theta)$.
Its bounded first partial derivatives on the safe set and the established bounds on $\zeta,\dot\zeta,\dot{\hat\theta}$ give a bounded derivative of $H(t,\hat\theta)\zeta$ almost everywhere by the chain rule.
It follows from the control law that $u$ is Lipschitz.
The first-derivative coefficient bounds, bounded $\dot z$, and bounded $\dot u^\star$ then make $\dot e$ uniformly continuous.
Since $e$ has a finite limit, this implies $\dot e\to0$: otherwise uniform continuity would give intervals of fixed length on which $\dot e$ keeps one sign and is bounded away from zero, contradicting convergence of $e$.
Finally, the error equation and control law give $u-u^\star=\frac{\dot e-\alpha e-\gamma z}{b}\longrightarrow0$ and $\widehat H\zeta-u^\star=(u-u^\star)+k\sigma_b e\longrightarrow0$.
This proves~\eqref{eq:r1-exact-convergence}.
\end{proof}

\subsection{Proof of Theorem~\ref{thm:model-family-regulation}}\label{app:model-family-regulation}
\begin{proof}
First consider how the common comparison bounds enter the analysis.
For a full transformed state, the Lyapunov construction in Appendix~\ref{app:fixed-model-regulation} provides one possible $W_K$.
The storage calculation, however, uses only the differential comparison inequality and the bound on $B_{e,K}\chi_K$ in~\eqref{eq:family-comparison-bounds}.
It therefore also applies to an effective comparison state retaining this coupling, without requiring a common positive definite matrix for the full stacked state.

We now prove the regulation claim using any $W_K$ satisfying that assumption.
The common constants give the same positive $c_e,\kappa$ in~\eqref{eq:r1-gain-conditions}.
Fix $K$ and an admissible initial condition.
The local solution argument in Supplementary~S1.3 is unchanged, since it uses the regularity of this fixed realization and its safe denominator, not a full-state lower bound for $W_K$.

Set $X_K=(W_K-M_{\chi e}\delta_K/\lambda_\chi)_+$ and $E_K=\mathcal D_{\delta_K}(e_K)$.
With $\tilde\theta_K=\hat\theta_K-\theta_K^\star$, use
\begin{equation*}
V_K=\frac{X_K^2}{2}+\frac{p_KE_K^2}{2}
+\frac{\tilde\theta_K^T\Gamma_K^{-1}\tilde\theta_K}{2}.
\end{equation*}
Local absolute continuity of $W_K$ justifies the squared positive-part calculation almost everywhere.
The matching identity~\eqref{eq:r1-matching} and the projection inequality give the same parameter cancellation as before.
Using~\eqref{eq:family-comparison-bounds} and~\eqref{eq:family-dead-zone} in the storage calculation yields
\begin{equation}\label{eq:family-storage-estimate}
\dot V_K\le-\frac{\lambda_\chi}{2}X_K^2-\frac{c_e}{2}E_K^2
+\frac{1}{2c_e}|\tau_K|^2.
\end{equation}
On the maximal existence interval, integration gives
\begin{equation}\label{eq:family-integrated-estimate}
\begin{aligned}
V_K(t)&+\int_{t_0}^t\left(\frac{\lambda_\chi}{2}X_K(s)^2
+\frac{c_e}{2}E_K(s)^2\right)ds\\
&\le V_K(t_0)+\frac{1}{2c_e}\|\tau_K\|_2^2.
\end{aligned}
\end{equation}
The right-hand side is finite for this $K$ and need not have a common upper bound as $K$ varies.
Since $p_K\ge p_{\min}$, it first bounds $e_K$.

It remains to recover states that an effective $W_K$ may not control.
The stable zero dynamics driven by the bounded $e_K$ bound $z$.
The full filter error satisfies
\begin{equation}\label{eq:family-full-filter-recovery}
\begin{aligned}
q_{f,K}&=\zeta_K-\zeta_K^\star-R_{f,K}e_K,\qquad R_{f,K}=G_K/b,\\
\dot q_{f,K}&=F_Kq_{f,K}-R_{f,K}\gamma z\\
&\quad+(F_KR_{f,K}-\dot R_{f,K}-R_{f,K}\alpha)e_K.
\end{aligned}
\end{equation}
For this fixed $K$, $F_K$ is uniformly exponentially stable and all forcing coefficients are bounded.
Variation of constants therefore bounds every component of $q_{f,K}$, including those omitted by the effective comparison state.
The first identity in~\eqref{eq:family-full-filter-recovery} then bounds $\zeta_K$ using its bounded comparison trajectory.
The safe denominator, compact candidate set, and fixed-realization coefficient bounds give bounded readout, regressor, input, and parameter update rate.
The ideal trajectory and inverse BI transformation recover the plant state as in Appendix~\ref{app:fixed-model-regulation}.

These state bounds are independent of the maximal existence time, so the local solution extends uniquely to $[t_0,\infty)$.
Neither common Euclidean bounds on the stacked states nor a common lower matrix bound for the full-state Lyapunov function is used.

Estimate~\eqref{eq:family-integrated-estimate} gives $E_K\in L^2$.
The established bounds make $\dot e_K$ bounded.
Thus $E_K$ is uniformly continuous and tends to zero by the square-integrability argument in Appendix~\ref{app:fixed-model-regulation}.
Since $|e_K|\le\delta_K+|E_K|$, the first assertion in~\eqref{eq:family-regulation} follows for every admissible trajectory.
No convergence assertion for $X_K$ is needed here.
Now let $K\to\infty$ and use~\eqref{eq:family-dead-zone} to obtain the stated iterated limit.
The same bound $\delta_K$ applies to each prescribed physical parameter and initial condition, but this argument does not interchange the time limit with a supremum over those data.
Finally, the choice $\delta_K=2\varepsilon_K/\kappa$ and $\varepsilon_K\to0$ give the practical regulation claim.
\end{proof}

\subsection{Proof of Lemma~\ref{lem:effective-realization-bounds}}
\label{app:effective-realization-bounds}
\begin{proof}
By~\eqref{eq:coefficient-c1-approx}, the comparison coefficients and their first time derivatives have bounds independent of $K$.
Equation~\eqref{eq:effective-forcing} involves only the fixed number $m$ of coefficient combinations.
It therefore gives common bounds on $\bar G_K,\dot{\bar G}_K$ and, since $|b|\ge b_{\min}$, on $\bar R_K,\dot{\bar R}_K$.
The comparison weights are constant.
Subtracting the effective filter equations and using $\bar R_Kb=\bar G_K$ cancels the input difference and gives~\eqref{eq:effective-cascade}.

The homogeneous cascade has uniformly exponentially stable diagonal systems and a uniformly bounded coupling.
Its exponential stability constants can therefore be chosen independently of $K$.
Applying the Lyapunov construction in Appendix~\ref{app:fixed-model-regulation} gives matrices $P_{\mathrm{eff},K}(t,\mu)$ with common positive scalar upper and lower bounds.
The resulting norm $W_K$ satisfies the required comparison inequality.
Together with~\eqref{eq:effective-matching} and the coefficient bounds, this proves~\eqref{eq:family-comparison-bounds}, with constants independent of the number of filter channels.

For each fixed $K$, initialize the ideal filter blocks at zero.
Stability of $F_0$ and bounded forcing give a bounded comparison trajectory.
Proposition~\ref{prop:forward-filter} separates the readout residual into coefficient mismatch and a finite-energy transient.
The convolution estimate~\eqref{eq:coefficient-to-residual} then gives~\eqref{eq:constructive-residual}.
Together with the safe candidate set, these properties verify Assumption~\ref{ass:finite-model} for each fixed $K$.
Actual filter initial values remain arbitrary.
Once the regulation estimate bounds $e_K$,~\eqref{eq:family-full-filter-recovery} bounds every actual filter state.
Explicit constants and the transient decomposition are given in Supplementary~S9.3.
\end{proof}

\section*{Acknowledgment}
This work was supported in part by the National Natural Science Foundation of China under Grants 52275564 and U24A20109.

\bibliographystyle{unsrtnat}
\bibliography{ref}

@article{davison1976robust,
  title={{The robust control of a servomechanism problem for linear time-invariant multivariable systems}},
  author={Davison, E. J.},
  journal={IEEE transactions on Automatic Control},
  volume={21},
  number={1},
  pages={25--34},
  year={1976},
  doi={10.1109/TAC.1976.1101137},
  publisher={IEEE},
}

@article{francis1976internal,
  title={{The internal model principle of control theory}},
  author={Francis, Bruce A and Wonham, W. M.},
  journal={Automatica},
  volume={12},
  number={5},
  pages={457--465},
  year={1976},
  doi={10.1016/0005-1098(76)90006-6},
  publisher={Elsevier},
}

@article{francis1977linear,
  title={{The linear multivariable regulator problem}},
  author={Francis, Bruce A},
  journal={SIAM Journal on Control and Optimization},
  volume={15},
  number={3},
  pages={486--505},
  year={1977},
  doi={10.1137/0315033},
  publisher={SIAM},
}

@article{isidori1990output,
  title={{Output regulation of nonlinear systems}},
  author={Isidori, A. and Byrnes, C. I.},
  journal={IEEE Transactions on Automatic Control},
  volume={35},
  number={2},
  pages={131--140},
  year={1990},
  doi={10.1109/9.45168},
  publisher={Institute of Electrical and Electronics Engineers (IEEE)},
}

@article{jiang2025output,
  title={{Output Feedback-Based Adaptive Optimal Output Regulation for Continuous-Time Strict-Feedback Nonlinear Systems}},
  author={Jiang, Yi and Chai, Tianyou and Chen, Guanrong},
  journal={IEEE Transactions on Automatic Control},
  year={2025},
  volume={70},
  number={2},
  pages={767--782},
  doi={10.1109/TAC.2024.3441668},
  publisher={IEEE},
}

@article{han2025iss,
  author={Han, Qi and Wang, Lei and Chen, Zhiyong},
  title={{An {ISS}-Based Design of Global Robust Adaptive Output Regulators for Nonlinear Systems}},
  journal={IEEE Transactions on Automatic Control},
  volume={70},
  number={5},
  pages={3354--3361},
  year={2025},
  doi={10.1109/TAC.2024.3512174},
}

@article{li2025global,
  title={{Global {$H_{\infty}$} Output Regulation of Nonlinear Systems Via Output-Feedback Reinforcement Learning}},
  author={Li, Shaobao and Wang, Yuxiang and Zhang, Yuguang and Luo, Xiaoyuan and Wang, Juan and Guan, Xinping},
  journal={IEEE Transactions on Automatic Control},
  year={2026},
  volume={71},
  number={3},
  pages={2107--2113},
  doi={10.1109/TAC.2025.3621992},
  publisher={IEEE},
}

@article{lin2025direct,
  title={{Direct Adaptive Cooperative Output Regulation of Unknown Multiagent Systems via Distributed Internal Model}},
  author={Lin, Liquan and Huang, Jie},
  journal={IEEE Transactions on Automatic Control},
  year={2025},
  volume={70},
  number={11},
  pages={7644--7651},
  doi={10.1109/TAC.2025.3576180},
  publisher={IEEE},
}

@article{gul2026multi,
  title={{Multi-partite output regulation of multi-agent systems}},
  author={G{\"u}l, K{\"u}r{\c{s}}ad Metehan and Sars{\i}lmaz, Selahattin Burak},
  journal={IEEE Transactions on Automatic Control},
  year={2026},
  volume={71},
  number={9},
  pages={6235--6242},
  doi={10.1109/TAC.2026.3682209},
  publisher={IEEE},
}

@article{li2026data,
  title={{Data-Driven Cooperative Output Regulation Via Output Feedback Control}},
  author={Li, Yifei and Liu, Wenjie and Wang, Gang and Xie, Lihua and Sun, Jian and Chen, Jie},
  journal={IEEE Transactions on Automatic Control},
  year={2026},
  volume={71},
  number={9},
  pages={6299--6306},
  doi={10.1109/TAC.2026.3682640},
  publisher={IEEE},
}

@article{pinzoni1993output,
  title={{Output regulation of linear time-varying systems}},
  author={Pinzoni, Stefano},
  journal={IFAC Proceedings Volumes},
  volume={26},
  number={2},
  pages={311--313},
  year={1993},
  doi={10.1016/S1474-6670(17)49133-6},
  publisher={Elsevier},
}

@article{zhang2006linear,
  title={{The linear periodic output regulation problem}},
  author={Zhang, Zhen and Serrani, Andrea},
  journal={Systems \& Control Letters},
  volume={55},
  number={7},
  pages={518--529},
  year={2006},
  doi={10.1016/j.sysconle.2005.11.002},
  publisher={Elsevier},
}

@article{ichikawa2006output,
  title={{Output regulation of time-varying systems}},
  author={Ichikawa, Akira and Katayama, Hitoshi},
  journal={Systems \& Control Letters},
  volume={55},
  number={12},
  pages={999--1005},
  year={2006},
  doi={10.1016/j.sysconle.2006.06.012},
  publisher={Elsevier},
}

@article{sun2009trajectory,
  title={{Trajectory tracking and disturbance rejection for linear time-varying systems: Input/output representation}},
  author={Sun, Zongxuan and Zhang, Zhen and Tsao, Tsu-Chin},
  journal={Systems \& Control Letters},
  volume={58},
  number={6},
  pages={452--460},
  year={2009},
  doi={10.1016/j.sysconle.2009.02.004},
  publisher={Elsevier},
}

@article{zhang2010novel,
  title={{A novel internal model-based tracking control for a class of linear time-varying systems}},
  author={Zhang, Zhen and Sun, Zongxuan},
  journal={Journal of Dynamic Systems, Measurement and Control, Transactions of the ASME},
  volume={132},
  number={1},
  pages={011004},
  year={2010},
  doi={10.1115/1.4000071},
  publisher={American Society of Mechanical Engineers (ASME)},
}

@article{zhang2014discrete,
  title={{A discrete time-varying internal model-based approach for high precision tracking of a multi-axis servo gantry}},
  author={Zhang, Zhen and Yan, Peng and Jiang, Huan and Ye, Peiqing},
  journal={ISA transactions},
  volume={53},
  number={5},
  pages={1695--1703},
  year={2014},
  doi={10.1016/j.isatra.2014.04.006},
  publisher={Elsevier},
}

@article{song2015low,
  title={{Low-order stabilizer design for discrete linear time-varying internal model-based system}},
  author={Song, Xingyong and Gillella, Pradeep Kumar and Sun, Zongxuan},
  journal={IEEE/ASME Transactions on Mechatronics},
  volume={20},
  number={6},
  pages={2666--2677},
  year={2015},
  doi={10.1109/TMECH.2015.2417125},
  publisher={IEEE},
}

@book{isidori2003robust,
  title={{Robust autonomous guidance: an internal model approach}},
  author={Isidori, Alberto and Marconi, Lorenzo and Serrani, Andrea},
  year={2003},
  doi={10.1007/978-1-4471-0011-9},
  publisher={Springer Science \& Business Media},
}

@ARTICLE{zhang2009tac,
  author={Zhang, Zhen and Serrani, Andrea},
  journal={IEEE Transactions on Automatic Control},
  title={{Adaptive Robust Output Regulation of Uncertain Linear Periodic Systems}},
  year={2009},
  volume={54},
  number={2},
  pages={266-278},
  doi={10.1109/TAC.2008.2010891},
}

@ARTICLE{marconi2013internal,
  author={Marconi, Lorenzo and Teel, Andrew R.},
  journal={IEEE Transactions on Automatic Control},
  title={{Internal Model Principle for Linear Systems With Periodic State Jumps}},
  year={2013},
  volume={58},
  number={11},
  pages={2788-2802},
  doi={10.1109/TAC.2013.2272137},
}

@article{carnevale2017robust,
  title={{Robust hybrid output regulation for linear systems with periodic jumps: Semiclassical internal model design}},
  author={Carnevale, Daniele and Galeani, Sergio and Menini, Laura and Sassano, Mario},
  journal={IEEE Transactions on Automatic Control},
  volume={62},
  number={12},
  pages={6649--6656},
  year={2017},
  doi={10.1109/TAC.2017.2716107},
  publisher={IEEE},
}

@article{zha2026robust,
  title={{Robust Output Regulation of Uncertain Linear Time-Varying Systems}},
  author={Zha, Jinmeng and Zhang, Zhen},
  journal={arXiv preprint arXiv:2601.17464},
  year={2026},
  doi={10.48550/arXiv.2601.17464},
}

@article{paunonen2017robust,
  title={{Robust output regulation for continuous-time periodic systems}},
  author={Paunonen, Lassi},
  journal={IEEE Transactions on Automatic Control},
  volume={62},
  number={9},
  pages={4363--4375},
  year={2017},
  doi={10.1109/TAC.2017.2654968},
  publisher={IEEE},
}

@article{bin2022robustness,
  title={{About robustness of control systems embedding an internal model}},
  author={Bin, Michelangelo and Astolfi, Daniele and Marconi, Lorenzo},
  journal={IEEE Transactions on Automatic Control},
  volume={68},
  number={3},
  pages={1306--1320},
  year={2023},
  doi={10.1109/TAC.2022.3151574},
  publisher={IEEE},
}

@article{byrnes2004nonlinear,
  author={Byrnes, Christopher I. and Isidori, Alberto},
  title={{Nonlinear Internal Models for Output Regulation}},
  journal={IEEE Transactions on Automatic Control},
  volume={49},
  number={12},
  pages={2244--2247},
  year={2004},
  doi={10.1109/TAC.2004.838492},
}

@article{forte2017robust,
  author={Forte, Francesco and Marconi, Lorenzo and Teel, Andrew R.},
  title={{Robust Nonlinear Regulation: Continuous-Time Internal Models and Hybrid Identifiers}},
  journal={IEEE Transactions on Automatic Control},
  volume={62},
  number={7},
  pages={3136--3151},
  year={2017},
  doi={10.1109/TAC.2016.2626143},
}

@article{bin2020class,
  author={Bin, Michelangelo and Marconi, Lorenzo},
  title={{{`Class-Type'} Identification-Based Internal Models in Multivariable Nonlinear Output Regulation}},
  journal={IEEE Transactions on Automatic Control},
  volume={65},
  number={10},
  pages={4369--4376},
  year={2020},
  doi={10.1109/TAC.2019.2955668},
}

@inproceedings{yang2010framework,
  author={Yang, Xi and Huang, Jie},
  title={{A Framework for Nonlinear Output Regulation for Time-varying Uncertain Systems}},
  booktitle={Proceedings of the 49th IEEE Conference on Decision and Control},
  year={2010},
  pages={5396--5401},
  doi={10.1109/CDC.2010.5717051},
}

@article{marconi2007output,
  author={Marconi, Lorenzo and Praly, Laurent and Isidori, Alberto},
  title={{Output Stabilization via Nonlinear {Luenberger} Observers}},
  journal={SIAM Journal on Control and Optimization},
  volume={45},
  number={6},
  pages={2277--2298},
  year={2007},
  doi={10.1137/050642344},
}

@article{bernard2020adaptive,
  author={Bernard, Pauline and Bin, Michelangelo and Marconi, Lorenzo},
  title={{Adaptive Output Regulation via Nonlinear {Luenberger} Observer-Based Internal Models and Continuous-Time Identifiers}},
  journal={Automatica},
  volume={122},
  pages={109261},
  year={2020},
  doi={10.1016/j.automatica.2020.109261},
}

@article{bin2020approximate,
  title={{Approximate nonlinear regulation via identification-based adaptive internal models}},
  author={Bin, Michelangelo and Bernard, Pauline and Marconi, Lorenzo},
  journal={IEEE Transactions on Automatic Control},
  volume={66},
  number={8},
  pages={3534--3549},
  year={2021},
  doi={10.1109/TAC.2020.3020563},
  publisher={IEEE},
}

@article{gentilini2023data,
  author={Gentilini, Lorenzo and Bin, Michelangelo and Marconi, Lorenzo},
  title={{Data-driven Output Regulation via {Gaussian} Processes and {Luenberger} Internal Models}},
  journal={IFAC-PapersOnLine},
  volume={56},
  number={1},
  pages={367--372},
  year={2023},
  doi={10.1016/j.ifacol.2023.02.062},
}

@article{han2025filter,
  author={Han, Qi and Wang, Lei and Marconi, Lorenzo and Zhang, Meng and Liu, Zhitao and Su, Hongye},
  title={{Adaptive Output Regulation with Internal Model-based Filter for Identification}},
  journal={IFAC-PapersOnLine},
  volume={59},
  number={19},
  pages={448--453},
  year={2025},
  doi={10.1016/j.ifacol.2025.11.075},
}

@article{wang2024nonparametric,
  title={{A nonparametric learning framework for nonlinear robust output regulation}},
  author={Wang, Shimin and Guay, Martin and Chen, Zhiyong and Braatz, Richard D},
  journal={IEEE Transactions on Automatic Control},
  volume={70},
  number={4},
  pages={2134--2149},
  year={2025},
  doi={10.1109/TAC.2024.3470065},
  publisher={IEEE},
}

\end{document}